\documentclass[journal,hidelinks]{IEEEtran}

\usepackage{amsmath,amsfonts,amssymb}
\usepackage{dsfont}
\usepackage{mathtools}
\usepackage{graphicx}
\usepackage{ifthen}
\usepackage{dblfloatfix}
\usepackage{placeins}
\usepackage{cite}
\usepackage{booktabs}
\usepackage{siunitx}
\usepackage[caption=false,font=normalsize,labelfont=sf,textfont=sf]{subfig}
\usepackage{algorithm}
\usepackage{algpseudocode}
\usepackage{orcidlink}
\usepackage{tikz}
\usepackage{upgreek}
\usepackage{xcolor}
\usetikzlibrary{arrows.meta,positioning,shapes.geometric}

\DeclareMathOperator*{\argmin}{arg\,min}

\newcommand{\rev}[1]{#1}
\newcommand{\revone}[1]{#1}
\newcommand{\revtwo}[1]{#1}
\newcommand{\revboth}[1]{#1}
\newcommand{\paperfig}[2][]{\includegraphics[#1]{#2}}

\begin{document}

\newcommand{\sampletick}[3]{\draw[#3, line width=0.9pt] (#1,#2) -- (#1,{#2+0.78});}

\title{Beyond the Doppler Dilemma: \\
Improved Fast Weather Radar Unambiguous Doppler Velocity Spectrum Reconstruction from Sparse Aperiodic Sweeps}

\author{Tworit Dash\,\orcidlink{0000-0003-3127-8841},~\IEEEmembership{Member, IEEE},
S.A.K. Syed Mohamed, Oleg Krasnov,~\IEEEmembership{Member, IEEE}, and Alexander Yarovoy,~\IEEEmembership{Fellow, IEEE}
\thanks{This work was partly financially supported by the Dutch Research Council (NWO) under contract WI-NC 175.2021.038.}
\thanks{Tworit Dash, Oleg Krasnov, and Alexander Yarovoy are with the Microwave Sensing, Signals and Systems (MS3) group at the Delft University of Technology, Delft, The Netherlands (e-mail: t.k.dash@tudelft.nl; o.a.krasnov@tudelft.nl; a.yarovoy@tudelft.nl).}
\thanks{S.A.K. Syed Mohamed was a MSc student at the Microwave Sensing, Signals and Systems (MS3) group at the Delft University of Technology (e-mail: S.A.K.SyedMohamed@student.tudelft.nl), now with Robin Radar Systems, The Hague, The Netherlands (e-mail: sultan.abdul@robinradar.com).}
}

\maketitle

\begin{abstract}
The problem of fast Doppler counter-aliasing for short-dwell weather-radar measurements is addressed. \revboth{The proposed approach realizes a log-periodic slow-time design by changing only the waiting time between otherwise unchanged frequency-modulated continuous-wave (FMCW) chirps.} The fast-time waveform, bandwidth, and range-processing chain are preserved while the acquisition changes the Doppler ambiguity structure. \revone{We formulate a novel moment-space ambiguity function for distributed weather spectra. It measures statistical ambiguity jointly in mean Doppler velocity and spectral width and expresses the competing sidelobes through their expected relative likelihood.} A complex Gaussian process (CGP) maximum-likelihood estimator reconstructs the spectrum from the measured aperiodic samples. \revone{A deterministic basin-selection and optimization strategy accelerates the same CGP likelihood without defining a different statistical estimator.} \revboth{Real precipitation measurements establish central-branch stability over an extended search domain.} \revtwo{Using 64 slow-time sweeps, we compare three acquisitions and their associated estimators. Periodic data are processed by the discrete Fourier transform (DFT) and parametric spectral estimation (PSE), log-periodic data by the nonuniform discrete Fourier transform (NUDFT) and CGP, and staggered data by velocity-difference and CGP estimators.} The differences become more pronounced when the dwell is reduced to 32 sweeps. \revtwo{The results show that aperiodic time sampling enables extended-domain branch identification without degrading spectral-width estimation.}

\end{abstract}


\section{Introduction}
\label{sec:introduction}
\IEEEPARstart{D}{oppler} weather radar measurements are normally interpreted within the maximum unambiguous velocity interval imposed by the Pulse Repetition Time (PRT). When the sampling in slow time is periodic, this interval is not only a property of the estimator but also a property of the measurement. The same complex time series is consistent with Doppler spectra shifted by integer multiples of the sampling frequency, and post-processing de-aliasing has to use spatial continuity, temporal continuity, additional scans, or external constraints to choose the correct branch \cite{James2001,Zhang2006,Gao2004,Louf2020}. In this sense, conventional Doppler processing often treats ambiguity after it has already been created by the acquisition.

\rev{This ambiguity is part of a broader radar-system bottleneck. A long operational range and high processing gain often require a long illumination time per measurement, for example a long FMCW chirp or a long phase-modulated continuous-wave (PMCW) code \cite{UtkuPMCW}. In such continuously transmitting radars, the waveform duration is tied to the maximum operational range and to the required processing gain, which is commonly associated with the time-bandwidth product of the target illumination, i.e., the product of the chirp or code duration seen by a target at the maximum range and the transmitted bandwidth. Such long fast-time waveforms are attractive for range sensitivity and range resolution, but with periodic triggering, they reduce the effective slow-time update rate and, therefore, the unambiguous Doppler interval. The resulting range-velocity coupling is particularly restrictive for FMCW and PMCW systems and is severe in weather-radar measurements, where operational ranges of dozens of kilometers must be combined with reliable reconstruction of the precipitation Doppler spectrum.} \revboth{The present realization addresses this bottleneck by changing the slow-time trigger locations through the waiting/reset times between otherwise unchanged FMCW chirps. The chirp duration, bandwidth, fast-time sampling, and range processing are retained, while the acquisition itself is given a less ambiguous Doppler response.}

Recent work on velocity recovery has also made important progress from another direction. The Doppler Velocity Recovery and De-aliasing algorithm (VRAD) blends long- and short-PRT scans to recover missing Doppler velocities in Weather Surveillance Radar-1988 Doppler split-cut observations \cite{Schvartzman2024}. Regression-enhanced processing further improves velocity retrievals in the Next Generation Weather Radar (NEXRAD) network by combining regression filtering with VRAD in the Sachidananda-Zrni\'c (SZ) phase-code processing chain configured as SZ(8/64) \cite{ImprovingNEXRADVelocityRetrievalsUsingMultiPRTScanswithRegressionProcessing}. These methods are designed for operational multi-PRT or phase-coded scan strategies and rely on information from different scans, censoring, filtering, and spatial continuity. They address ambiguity recovery after or across scans, whereas the problem considered here is whether one short high-precision aperiodic dwell can already support unambiguous spectral inversion over an extended Doppler domain before any spatial de-aliasing step is introduced.

The novel counter-aliasing view adopted here is therefore a measurement-design view. Instead of first creating an aliased measurement and then resolving it, the slow-time samples can be placed aperiodically so that the measurement itself spans a much larger Doppler interval \cite{Dash2024Counter-AliasingTrain}. For a finite aperiodic sequence, the relevant theoretical interval is determined by the finest time grid on which the samples are represented. Therefore, if the time stamps are controlled or recorded with high precision, the formal unambiguous interval can be much larger than the Nyquist interval of the corresponding periodic sequence. This makes it possible to regard ambiguity prevention as a measurement-design problem rather than only a post-processing problem.

\revboth{Our earlier work \cite{Dash2024Counter-AliasingTrain} proposes the counter-aliasing principle and the analytical log-periodic waiting-time rule. Its simulations establish that a CGP model can retrieve a sufficiently narrow weather-like Doppler spectrum outside the periodic Nyquist interval and identify real-radar realization as the next step. The present paper carries out that realization on an FMCW weather radar using the measured high-precision timestamps.} \rev{Although the estimator is developed here for precipitation, the acquisition principle is not tied to precipitation alone. Other radar targets could be treated within the same slow-time counter-aliasing framework if a suitable stochastic or parametric covariance model is available for their Doppler response.}

A central advantage of the proposed acquisition is that counter-aliasing is achieved by programming the waiting/reset times between successive FMCW chirps while preserving the fast-time chirp, bandwidth, and range-processing chain. The Doppler processing is then performed directly from the measured slow-time samples. \revboth{This experimental realization moves the existing timing law from simulation to measured precipitation echoes.} \revone{For practical inversion, we develop a deterministic basin-selection strategy for the existing CGP maximum-likelihood estimator. It preserves the statistical model and objective function of the full multi-start estimator, but replaces its random starting points by structured candidates, successive likelihood refinement, and one final local optimization.}

The novelty of this work is \rev{fourfold}.
\begin{enumerate}
    \item \revboth{The log-periodic waiting-time design proposed in \cite{Dash2024Counter-AliasingTrain} is realized on an FMCW weather radar. Only the slow-time triggering pattern is changed; the fast-time chirp and range-processing chain remain fixed.}
    \item \revone{A moment-space ambiguity function is formulated for extended stochastic Doppler spectra. Unlike a point-target ambiguity function, it quantifies ambiguity jointly in mean Doppler velocity and spectral width from the sampled covariance and provides the main-lobe width and ambiguity-sidelobe levels of each timing law in physically interpretable moment coordinates.}
    \item \revone{A deterministic basin-selection strategy is developed for the CGP maximum-likelihood estimator. Structured candidates obtained from NUDFT peaks and ambiguity translates are ranked and refined before one final optimization. This accelerates the existing estimator without presenting the search strategy as a new statistical estimator.}
    \item \revboth{The acquisition and inversion are evaluated in two complementary ways. Measured precipitation data quantify central-branch stability under periodic, staggered, and log-periodic timing, while controlled phase modulation of the same complex echoes places the spectrum on known Doppler branches and permits direct error assessment.}
\end{enumerate}

Taken together, these contributions show that preventing Doppler aliasing through deliberate aperiodic slow-time design is not only theoretically appealing but also experimentally realizable, and that the associated inversion can be made fast enough to support practical deployment.

The main body of the paper is organized as follows. Section~\ref{sec:related_work} reviews the relevant literature and motivates the present counter-aliasing perspective. Section~\ref{sec:signal_model} introduces the signal model and noise calibration. Section~\ref{sec:sampling_precision} defines the normalized Doppler quantities, the measured sampling sequences, and the corresponding theoretical ambiguity intervals. Section~\ref{sec:extended_ambiguity} introduces the extended-spectrum ambiguity analysis used to compare sampling sequences before the real-data evaluation. Section~\ref{sec:estimators} describes \revtwo{the Fourier, velocity-difference, and covariance-based estimators}, and Section~\ref{sec:spectrum_reconstruction} summarizes the frequency-domain spectrum reconstruction. Section~\ref{sec:real_data} presents the real-data analysis, measurement setup, and preprocessing steps. The experimental results are reported in Section~\ref{sec:results}, followed by the computational complexity discussion in Section~\ref{sec:complexity}, the interpretation of the findings in Section~\ref{sec:discussion}, and the concluding remarks in Section~\ref{sec:conclusion}.

\section{Related Work and Rationale}
\label{sec:related_work}
Spectral moment estimation in weather radar is commonly based on the periodogram, pulse-pair estimators, or parametric covariance models \cite{Doviak1993,Zrnic1975VelocityRadar}. For uniformly sampled data, the discrete Fourier transform (DFT) is computationally efficient and directly connected to the PRT. Its limitation is also direct: the Doppler spectrum is periodic in frequency, and the velocity estimate is meaningful only within the principal unambiguous interval unless an external de-aliasing step is used.

The ambiguity problem has usually been handled after acquisition. Operational de-aliasing methods use continuity in range, azimuth, elevation, or time to choose one of the periodic velocity branches \cite{James2001,Zhang2006,Louf2020}. Variational and wind-retrieval methods can also use physical constraints to recover velocity or wind fields from aliased radial velocity observations \cite{Gao2004}. These methods are effective when the weather field is spatially coherent and sufficient context is available, but they do not change the information content of a single dwell. The single-cell ambiguity remains present, and the de-aliasing step becomes a separate inference problem. This distinction is important for short dwell processing, phased-array radar updates, and vertically pointing observations, where the estimator may be required to work with limited spatial context.

\revtwo{Staggered PRT is a sampling strategy that alternates two pulse repetition intervals. A common velocity-difference estimator combines the corresponding pulse-pair estimates to increase the effective unambiguous velocity \cite{Zrnic1977,Sachidananda2000ClutterPRT,Torres2004DesignWSR-88D}. Staggered acquisitions and their associated estimators have also been developed for moment estimation, range-overlay mitigation, and clutter filtering in weather radar \cite{Sachidananda2000ClutterPRT,Sachidananda2003UnambiguousTechnique,Moisseev2008ClutterWaveforms,Warde2017Staggered-PRTFilter,Torres2017}. They have therefore been very influential in operational radar processing.} However, the staggered sequence is still periodic on a larger hidden grid. If the hidden integer periods are large, the theoretical unambiguous interval can be large, but the ambiguity transfer function becomes sensitive to noise and spectral width. Therefore, the theoretical interval and the practical reliability of branch selection are not the same quantity.

More recent operational work uses multiple scans or multiple PRTs to recover velocities that are missing or aliased in one scan. VRAD, the Doppler Velocity Recovery and Dealiasing algorithm, uses long-PRT surveillance-scan velocities to populate obscured regions and to assist dealiasing of short-PRT Doppler-scan velocities \cite{Schvartzman2024}. Regression-enhanced VRAD combines this recovery framework with regression processing of SZ(8/64) data to reduce velocity voids in the Next Generation Weather Radar (NEXRAD) network \cite{ImprovingNEXRADVelocityRetrievalsUsingMultiPRTScanswithRegressionProcessing}. These approaches address an important operational range-velocity problem, especially in low-elevation surveillance where range overlays and ``purple haze'' are central concerns. Their logic is nevertheless different from the one used here. They improve the recovery of a velocity field by combining information from different scans and by enforcing spatial consistency. The present approach instead tests whether the sampled signal from one deliberately aperiodic dwell already contains enough information to estimate the Doppler spectrum on an extended velocity axis.

Nonuniform Fourier methods, including the direct transform (i.e., NUDFT) and the Lomb-Scargle periodogram, allow spectral analysis at arbitrary sample times \cite{Scargle1982StudiesData,VanderPlas2018UnderstandingPeriodogram}. They are attractive because they do not require resampling the time series onto a uniform grid. \revtwo{The DFT evaluates the Fourier sum on the uniformly sampled periodic acquisition, whereas the NUDFT evaluates it at the nonuniform log-periodic timestamps.} The same family of ideas appears in nonuniform PRT radar processing and random sampling theory, where the goal is to reduce coherent grating responses by modifying the sampling pattern \cite{Doerry2017RadarPRF,Gaster1977TheTransform}. However, the periodogram is a quadratic transform of the finite observation window. For sparse short dwells, the sidelobe structure of the sampling pattern can be comparable with the weather echo power. For a distributed target with a nonzero Doppler width, these artifacts may bias the spectral moments when the integration interval is extended \cite{S.A.KSyedMohamed2024ApplicationSpectrum}.

There is also a broad literature on improved spectral estimation beyond the simple periodogram. Autoregressive and adaptive estimators have been used for atmospheric Doppler radar and weather radar spectral moments \cite{Palmer1991,Pinsky2011,Warde2014TheAnalysis}. Multitaper and other small-sample spectral estimators reduce variance by controlling the spectral window \cite{Thomson1982SpectrumAnalysis,Das2021SpectralApproach}. Sparse and compressive approaches can exploit compact spectral support, but their performance depends on the adequacy of the sparsifying model and the dictionary used for reconstruction \cite{Ariananda2012CompressiveEstimation,Purnamasari2021}. These methods are useful references, but precipitation Doppler spectra are not single sinusoids. They are distributed weather echoes with physically meaningful width, and the estimator must preserve both the mean velocity and the spectral width.

\revtwo{Finite-dwell spectral broadening is especially important when the weather spectrum is narrower than the Fourier resolution. Here, the finite dwell is the limited observation time used to form one Doppler estimate. The parametric semianalytical-model-based spectral estimator (PSE) in \cite{Dash2024DopplerModel} uses a Whittle likelihood and explicitly includes this finite observation window in its expected periodogram model. It therefore provides an intermediate reference between uncorrected Fourier moments, which contain the window convolution, and the time-domain CGP likelihood, which is formed directly from the measured samples. Here the PSE is applied both to uniformly sampled simulations and to the measured periodic acquisition. Its existing uniform-time formulation is not applied to the log-periodic samples.}

Bayesian and Gaussian-process (GP) spectral approaches provide another route. A stationary covariance function in time has a corresponding power spectral density, and the frequency-domain posterior can be evaluated directly from the time-domain observations \cite{Tobar2018BayesianEstimation,Dash2023PrecipitationPrior}. Complex-valued Gaussian processes (CGP) provide the natural probabilistic setting for complex radar echoes \cite{Boloix-Tortosa2018ComplexRegression}. In weather radar, a Gaussian-shaped Doppler spectrum gives a covariance that is a complex sinusoid multiplied by a Gaussian envelope. This parameterization is compact, physically interpretable, and well-suited to short dwell processing. It does not interpolate the signal in slow time (which is the focus of most GP-based literature, and which usually does not address the ambiguity issue); instead, it directly reconstructs the high-resolution spectrum using only the short-dwell data. We introduce a novel integrative approach that combines covariance-based inversion with high-precision aperiodic slow-time sampling and a deterministic search strategy that avoids large random multi-start optimization.

\section{Signal Model and Noise Calibration}
\label{sec:signal_model}
For each range cell, the slow-time precipitation signal is represented by the observation vector
\begin{equation}
    \mathbf{z}=[z(t_0),z(t_1),\ldots,z(t_{N-1})]^T,
\end{equation}
where $N=64$ unless otherwise stated and $t_k$ denotes slow time normalized by the first sampling interval. The measurement model is
\begin{equation}
    \mathbf{z}=\mathbf{s}+\mathbf{n},\qquad
    n_k\sim\mathcal{CN}(0,\sigma_n^2).
    \label{eq:measurement_model}
\end{equation}
The weather signal is modeled as a zero-mean proper CGP with covariance
\begin{equation}
    C_\theta(\tau)
    =P\exp(-2\uppi^2\sigma_{fn}^2\tau^2)\exp(j2\uppi\mu_{fn}\tau),
    \label{eq:single_covariance}
\end{equation}
where $\tau=t_a-t_b$, $P$ is the signal power, $\mu_{fn}$ is the normalized mean Doppler frequency, and $\sigma_{fn}$ is the normalized Doppler spectral width. Following the complex Gaussian process weather-spectrum covariance model used for Doppler spectrum reconstruction in \cite{Dash2023PrecipitationPrior,Tobar2018BayesianEstimation}, the full covariance matrix is
\begin{equation}
    \mathbf{K}_\theta
    =
    \left[C_\theta(t_a-t_b)\right]_{a,b=0}^{N-1}
    +\sigma_n^2\mathbf{I}.
    \label{eq:covariance_matrix}
\end{equation}
\revtwo{For periodic sampling, the covariance depends only on the integer lag and $\mathbf{K}_\theta$ is Toeplitz. Aperiodic sampling produces a general Hermitian covariance matrix because equal index separations no longer imply equal time lags. This change in the pairwise-lag structure breaks the exact periodic repetition in mean Doppler frequency and requires the dense factorization used in the likelihood evaluation.}
The negative log-likelihood used for inversion is
\begin{equation}
    \mathcal{L}(\theta)
    =
    \log|\mathbf{K}_\theta|
    +
    \mathbf{z}^{H}\mathbf{K}_\theta^{-1}\mathbf{z}
    + N\log\uppi,
    \label{eq:nll}
\end{equation}
where $\theta=[\mu_{fn},\sigma_{fn}]^T$. The signal power $P$ is treated as a nuisance parameter rather than as an optimized element of $\theta$: for each range cell and dwell, it is fixed from the average range fast Fourier transform (FFT) power before optimizing the Doppler mean and width. The noise variance $\sigma_n^2$ is also fixed from calibration. The power can be estimated jointly with the Doppler parameters, and the multiple-target extension in \cite{Dash2024Counter-AliasingTrain} uses this broader view; fixing $P$ keeps the real-data inversion focused on aliasing and computation. The noise is estimated from a separate noise-only measurement acquired with periodic slow-time sampling. The same fast-time preprocessing and range FFT are applied to the noise-only data; for each range cell, the noise power is averaged over slow time, the median over available noise bursts is taken, and a moving median over range is used to suppress isolated calibration fluctuations. The resulting $\sigma_n^2$ is kept fixed for all estimators.

Branch denotes one repeated Doppler interval that can look plausible under ambiguous sampling, similar to one folded velocity interval in radar. Basin denotes the region of the likelihood surface that pulls an optimizer toward the same local solution. A wrong branch therefore corresponds to a physically wrong extended-domain velocity interval, whereas a wrong basin is the optimization manifestation of that ambiguity. The physical interpretation of the retrieved moments is restricted to range cells containing precipitation echo power above the calibrated noise background. In weak-echo or clutter-contaminated cells, the likelihood over the extended Doppler interval can become broad or multimodal. The multimodal Doppler-spectrum case is treated in \cite{Dash2024Counter-AliasingTrain}; the present analysis focuses on the real-time implementation and real-radar realization of counter-aliasing for one extended weather target.

\section{Sampling Precision and Doppler Windows}
\label{sec:sampling_precision}
\subsection{Normalized Doppler Frequency}
Let $T_0$ denote the first slow-time interval, which is \SI{0.5}{ms} in the measurements considered in Section~\ref{sec:real_data}. The normalized Doppler frequency is
\begin{equation}
    f_n = f_D T_0 = \frac{2vT_0}{\lambda},
    \label{eq:normalized_frequency}
\end{equation}
where $f_D$ is the Doppler frequency, $v$ is the radial velocity, and $\lambda$ is the wavelength. For a \SI{3.315}{GHz} radar carrier, $\lambda \approx \SI{9.05}{cm}$. A periodic sequence with interval $T_0$ has the principal normalized interval
\begin{equation}
    f_n \in \left[-\frac{1}{2},\frac{1}{2}\right],
    \qquad
    v \in \left[-\frac{\lambda}{4T_0},\frac{\lambda}{4T_0}\right].
    \label{eq:periodic_nyquist}
\end{equation}

For an aperiodic sequence represented on a hidden clock grid, let the absolute slow-time sample locations be
\begin{equation}
    \tilde{t}_k = q_k T_c,\qquad q_k \in \mathbb{Z},
    \label{eq:clock_grid}
\end{equation}
where $T_c$ is the timestamp clock period. For the timing headers used in the real-data analysis, the integer timestamp represents the number of \SI{120}{MHz} clock cycles separating the first acquisition trigger from the trigger of the snapshot of interest. One timing tick is therefore $T_c=1/(120\times10^6)\,\mathrm{s}\approx\SI{8.33}{ns}$. The estimators use normalized sample locations $t_k=(\tilde{t}_k-\tilde{t}_0)/T_0=(q_k-q_0)/(q_1-q_0)$, so the first interval is one normalized unit.
If the greatest common divisor of all slow-time increments is $g$ clock cycles, the finest repeated time unit in the observed sequence is $T_u=gT_c$. The corresponding theoretical normalized Doppler fold is
\begin{equation}
    F_{\mathrm{fold}} = \frac{T_0}{T_u}
    = \frac{(q_1-q_0)T_c}{gT_c}
    = \frac{q_1-q_0}{g},
    \label{eq:clock_fold}
\end{equation}
and the theoretical interval is $f_n \in [-F_{\mathrm{fold}}/2,F_{\mathrm{fold}}/2]$. For both measured aperiodic timing sequences, the integer slow-time increments have greatest common divisor
\begin{equation}
    g=\gcd(\Delta q_1,\Delta q_2,\ldots)=1 .
\end{equation}
Thus the finest hidden-grid time unit is $T_u=gT_c=T_c\approx\SI{8.33}{ns}$. The formal normalized clock-grid interval follows from the measured integer timestamps through \eqref{eq:clock_fold}.
The practical search interval is chosen as
\begin{equation}
    \mathcal{F}_{n_f} = \left[-\frac{n_f}{2},\frac{n_f}{2}\right],
    \qquad n_f=16,
    \label{eq:processing_window}
\end{equation}
\revtwo{which is much larger than the periodic Nyquist interval while remaining a controlled interval for comparing the sampling strategies and their associated estimators.}

\subsection{Sampling Sequences}
Three measured slow-time sequences are considered. The \textbf{periodic sequence} is modeled as
\begin{equation}
    t_k^{(\mathrm{per})}=k,\qquad k=0,\ldots,N-1,
    \label{eq:periodic_sequence}
\end{equation}
after normalization by its minimum interval $T_0=\SI{0.5}{ms}$. Hence, the principal periodic Nyquist interval is $f_n\in[-0.5,0.5]$, which corresponds to the physical velocity interval
\begin{equation}
    |v| \le \frac{\lambda}{4T_0} \approx V_{a}^{( \mathrm{per})} = \SI{45.25}{m/s}.
\end{equation}

The \textbf{log-periodic sequence} follows the design proposed in \cite{Dash2024Counter-AliasingTrain}, but the real-data inversion uses the exact recorded timestamps rather than the rounded analytical positions. \revboth{It is a tunable aperiodic timing law. The curvature parameter $d_2$ controls the progressive departure from a uniform grid and thereby changes the distribution of pairwise lags that determines the moment-space ambiguity function. The limiting case $d_2\rightarrow0$ is periodic sampling, whereas increasing $d_2$ spreads the ambiguity sidelobes at the cost of a longer dwell when the minimum interval and sample count are fixed. Fig.~\ref{fig:ambiguity_design} marks the implemented value within this trade-off.} The symbol $N_{\mathrm{ap}}$ denotes the number of aperiodic slow-time samples in the designed sequence. If a periodic and an aperiodic dwell are forced to occupy the same total time duration, $N_{\mathrm{ap}}$ can be smaller than the periodic sample count $N$ because the aperiodic intervals may become longer. The measured comparisons use the recorded 64-sweep dwells and their exact header timestamps. The normalized sequence abides by the following rule:
\begin{equation}
   t_k^{(\mathrm{log})} = \frac{d_1}{T_0 d_2} \left[\exp\left(d_2k\right) - 1\right], \quad k = 0, 1, 2, \dots, N_{\mathrm{ap}}-1,
\end{equation} 
where $d_2$ controls the non-linearity in the sequence (set to 0.007423 for the measured acquisition), and $d_1$ controls the minimum sampling interval (i.e., here the same as $T_0$).
Let $c_k^{(\mathrm{log})}$ denote the recorded header time stamps in clock cycles, and let $\Delta c_{\min}^{(\mathrm{log})}=59792$ be the measured minimum increment in the considered dwell. The normalized slow-time grid used by the estimators is
\begin{equation}
    t_k^{(\mathrm{log})}
    = \frac{c_k^{(\mathrm{log})}-c_0^{(\mathrm{log})}}{\Delta c_{\min}^{(\mathrm{log})}},
    \label{eq:log_sequence}
\end{equation}
so that the first measured points are
\begin{equation}
    t_k^{(\mathrm{log})}
    = [0,\ 1,\ 2.0074,\ 3.0222,\ 4.0448,\ 5.0748,\ \ldots ].
\end{equation}

The \textbf{staggered sequence} also uses the measured header time stamps, normalized by the measured short interval $\Delta c_{\min}^{(\mathrm{stag})}=59566$ clock cycles,
\begin{equation}
    t_k^{(\mathrm{stag})}
    = \frac{c_k^{(\mathrm{stag})}-c_0^{(\mathrm{stag})}}{\Delta c_{\min}^{(\mathrm{stag})}}.
\end{equation}
Thus, the first measured points are
\begin{equation}
    t_k^{(\mathrm{stag})}
    = [0,\ 1,\ 2.5002,\ 3.5002,\ 5.0004,\ 6.0004,\ \ldots ].
\end{equation}
The measured short and long intervals are represented as integer clock cycles. For the first pair in the considered dwell,
\begin{equation}
    \frac{\Delta t_1}{\Delta t_2}
    = \frac{59566}{89359}
    = 0.666592061236,
    \label{eq:staggered_ratio}
\end{equation}
which is close to, but not exactly, $2/3$. This distinction is important because the theoretical ambiguity interval follows the actual clock-grid representation rather than the rounded ratio.

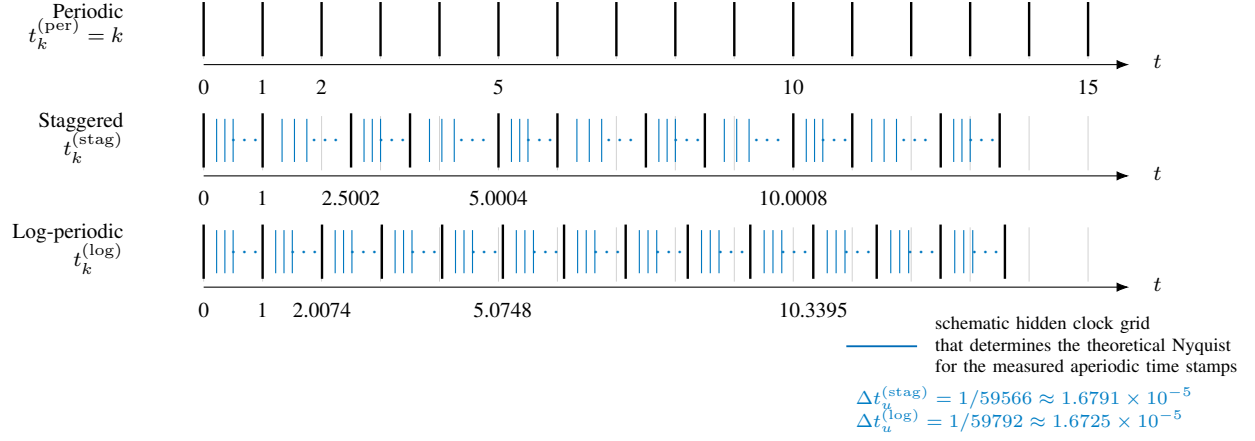
\begin{figure*}[!t]
\centering
\begin{tikzpicture}[
    x=0.78cm,
    y=0.92cm,
    every node/.style={font=\footnotesize}
]
\node[anchor=east, align=right] at (-1.2,0.55) {Periodic\\$t_k^{(\mathrm{per})}=k$};
\draw[-Latex] (0,0) -- (15.7,0);
\foreach \x in {0,...,15} {
    \draw[gray!35, line width=0.3pt] (\x,0.08) -- (\x,0.86);
}
\foreach \x in {0,1,2,3,4,5,6,7,8,9,10,11,12,13,14,15} {
    \sampletick{\x}{0.12}{black}
}
\foreach \x/\lab in {0/0,1/1,2/2,5/5,10/10,15/15} {
    \node[anchor=north] at (\x,-0.08) {\lab};
}
\node[anchor=west] at (15.95,0.04) {$t$};

\node[anchor=east, align=right] at (-1.2,-1.05) {Staggered\\$t_k^{(\mathrm{stag})}$};
\draw[-Latex] (0,-1.6) -- (15.7,-1.6);
\foreach \x in {0,...,15} {
    \draw[gray!35, line width=0.3pt] (\x,-1.52) -- (\x,-0.74);
}
\foreach \x in {0,1,2.5002,3.5002,5.0004,6.0004,7.5006,8.5006,10.0008,11.0008,12.5010,13.5010} {
    \sampletick{\x}{-1.48}{black}
}
\foreach \x/\lab in {0/0,1/1,2.5002/2.5002,5.0004/5.0004,10.0008/10.0008} {
    \node[anchor=north] at (\x,-1.68) {\lab};
}
\foreach \a/\b in {0/1,1/2.5002,2.5002/3.5002,3.5002/5.0004,5.0004/6.0004,6.0004/7.5006,7.5006/8.5006,8.5006/10.0008,10.0008/11.0008,11.0008/12.5010,12.5010/13.5010} {
    \pgfmathsetmacro{\xone}{\a + 0.22*(\b-\a)}
    \pgfmathsetmacro{\xtwo}{\a + 0.36*(\b-\a)}
    \pgfmathsetmacro{\xthr}{\a + 0.50*(\b-\a)}
    \pgfmathsetmacro{\xmid}{\a + 0.73*(\b-\a)}
    \draw[cyan!55!blue, line width=0.42pt] (\xone,-1.40) -- (\xone,-0.78);
    \draw[cyan!55!blue, line width=0.42pt] (\xtwo,-1.40) -- (\xtwo,-0.78);
    \draw[cyan!55!blue, line width=0.42pt] (\xthr,-1.40) -- (\xthr,-0.78);
    \node[text=cyan!55!blue, font=\small] at (\xmid,-1.10) {$\cdots$};
}
\node[anchor=west] at (15.95,-1.56) {$t$};

\node[anchor=east, align=right] at (-1.2,-2.65) {Log-periodic\\$t_k^{(\mathrm{log})}$};
\draw[-Latex] (0,-3.2) -- (15.7,-3.2);
\foreach \x in {0,...,15} {
    \draw[gray!35, line width=0.3pt] (\x,-3.12) -- (\x,-2.34);
}
\foreach \x in {0,1,2.0074,3.0222,4.0448,5.0748,6.1125,7.1577,8.2107,9.2712,10.3395,11.4152,12.4986,13.5896} {
    \sampletick{\x}{-3.08}{black}
}
\foreach \x/\lab in {0/0,1/1,2.0074/2.0074,5.0748/5.0748,10.3395/10.3395} {
    \node[anchor=north] at (\x,-3.28) {\lab};
}
\foreach \a/\b in {0/1,1/2.0074,2.0074/3.0222,3.0222/4.0448,4.0448/5.0748,5.0748/6.1125,6.1125/7.1577,7.1577/8.2107,8.2107/9.2712,9.2712/10.3395,10.3395/11.4152,11.4152/12.4986,12.4986/13.5896} {
    \pgfmathsetmacro{\xone}{\a + 0.22*(\b-\a)}
    \pgfmathsetmacro{\xtwo}{\a + 0.36*(\b-\a)}
    \pgfmathsetmacro{\xthr}{\a + 0.50*(\b-\a)}
    \pgfmathsetmacro{\xmid}{\a + 0.73*(\b-\a)}
    \draw[cyan!55!blue, line width=0.42pt] (\xone,-3.00) -- (\xone,-2.38);
    \draw[cyan!55!blue, line width=0.42pt] (\xtwo,-3.00) -- (\xtwo,-2.38);
    \draw[cyan!55!blue, line width=0.42pt] (\xthr,-3.00) -- (\xthr,-2.38);
    \node[text=cyan!55!blue, font=\small] at (\xmid,-2.70) {$\cdots$};
}
\node[anchor=west] at (15.95,-3.16) {$t$};

\draw[cyan!45!blue, line width=0.7pt] (10.9,-4.05) -- (12.1,-4.05);
\node[anchor=west, align=left, font=\scriptsize] at (12.25,-4.05)
{schematic hidden clock grid\\that determines the theoretical Nyquist\\for the measured aperiodic time stamps};
\node[anchor=west, align=left, font=\scriptsize, text=cyan!60!blue] at (10.9,-4.95)
{$\Delta t_u^{(\mathrm{stag})}=1/59566\approx 1.6791\times 10^{-5}$\\
$\Delta t_u^{(\mathrm{log})}=1/59792\approx 1.6725\times 10^{-5}$};
\end{tikzpicture}
\caption{Schematic of the normalized slow-time sampling grids used in the paper. The three rows share comparable normalized time axes. The periodic sequence is displayed on the unit grid defined by the common reference interval $T_0$. The staggered and log-periodic sequences are drawn from the measured header time stamps after normalization by their respective minimum intervals. The blue auxiliary marks represent hidden clock-grid locations between adjacent samples; three marks and an ellipsis schematically represent the much denser underlying grid. For the displayed normalization, one hidden-grid step corresponds to $\Delta t_u^{(\mathrm{stag})}=1/59566\approx 1.6791\times 10^{-5}$ in the staggered case and $\Delta t_u^{(\mathrm{log})}=1/59792\approx 1.6725\times 10^{-5}$ in the log-periodic case. This underlying grid determines the formal theoretical Nyquist interval and is not shown to scale.}
\label{fig:sampling_schematic}
\end{figure*}

Fig.~\ref{fig:sampling_schematic} gives a schematic view of the three normalized slow-time grids used throughout the paper. The periodic case lies on a uniform unit-spaced grid, the staggered case alternates short and long separations on the same normalized axis, and the log-periodic case follows the measured progressively perturbed placement. For all three acquisitions the fast-time FMCW chirp is the same; the proposed counter-aliasing action is to change only the wait/reset time between chirps, which changes the slow-time sampling grid without changing the chirp waveform, chirp bandwidth, or range processing. \rev{This is the practical mechanism by which the experiment modifies the Doppler ambiguity structure while preserving the range waveform and its processing gain.}

All three sequences have the same minimum sampling interval, and no measured sample is available on a finer slow-time grid. This gives a fair comparison among the timing strategies. 

\section{Moment-Space Ambiguity Function for Extended Doppler Spectra}
\label{sec:extended_ambiguity}
\revone{The clock-grid argument in Section~\ref{sec:sampling_precision} defines where exact repetition can occur. In this section, we formulate a moment-space ambiguity function whose coordinates are the mean Doppler frequency and Doppler spectral width. This extends the familiar point-target ambiguity concept from one frequency coordinate to the two extended weather-spectrum moments that govern the covariance of a Gaussian precipitation echo.  This novel ambiguity function is an informative and effective way to assess the performance of sampling strategies for random processes, as it models their covariance with a lower-dimensional representation (explainable by a few variables; here, those are the statistical spectral moments).} To visualize the difference between the ambiguity function of a deterministic point frequency-based signal and a random process, let's consider a conventional slow-time ambiguity or spectral-window analysis that considers the response of a point Doppler component,
\begin{equation}
    A_{\mathrm{point}}(\Delta f)
    =
    \left|
    \sum_{k=0}^{N-1}\exp(-j2\uppi\Delta f t_k)
    \right|^2.
    \label{eq:point_ambiguity}
\end{equation}
This mask response describes sidelobes produced by the sampling pattern. However, precipitation is an extended stochastic Doppler spectrum rather than a single point frequency. \revtwo{In meteorological processing of staggered PRT data, ambiguity is often described through branch selection of the mean Doppler velocity.} That view is also incomplete for the present problem because the Doppler spectrum width changes the time covariance and therefore changes which slow-time lags are informative. \revone{In moment space, the main lobe is the neighborhood of the true mean-width pair. A sidelobe is a separated candidate pair that produces a similar sampled covariance and therefore remains undesirably plausible. Mean-direction sidelobes correspond to competing velocity branches, while displacement in the width direction indicates ambiguity or weak resolution in spectral width.}

\revone{To compare periodic, staggered, and log-periodic sampling for extended weather targets, we define the moment-space ambiguity function between complete Doppler spectra.} Consider the single-component Gaussian Doppler spectrum
\begin{equation}
    \begin{aligned}
    S_\theta(f)
    &=
    \frac{P}{\sqrt{2\uppi\sigma_{fn}^2}}
    \exp\left[-\frac{(f-\mu_{fn})^2}{2\sigma_{fn}^2}\right],\\[-2pt]
    \theta&=[\mu_{fn},\sigma_{fn}]^T,
    \end{aligned}
    \label{eq:ambiguity_spectrum}
\end{equation}
where $\mu_{fn} = \mu_{v}/(2V_{a}^{( \mathrm{per})})$ is the normalized mean Doppler frequency, and $\sigma_{fn} = \sigma_{v}/(2V_{a,\mathrm{per}})$ is the normalized Doppler spectrum width.
As in the signal model, $P$ is treated as a nuisance scale parameter and is kept fixed when comparing ambiguity over $\theta$. This divergence analysis does not fit a periodogram, posterior spectrum, or optimizer output; it evaluates whether two candidate Gaussian weather spectra would produce distinguishable sampled covariance matrices.
\revboth{Because this function is evaluated before any record is generated or inverted, it characterizes the timing law and covariance family independently of the numerical estimator. Its central-lobe extent measures local joint resolution in mean Doppler and spectral width, while its remote sidelobe levels measure the relative plausibility of competing mean--width solutions.}
\revtwo{The ambiguity functions and false-detection calculations use $P=1$ and $\sigma_n^2=0.01$, corresponding to an input signal-to-noise ratio (SNR) of \SI{20}{dB}. For the radar parameters in Section~\ref{sec:real_data}, the physical width is $\sigma_v=90.5\sigma_{fn}$~\si{m.s^{-1}}; the displayed normalized interval $[0,0.20]$ therefore spans physical widths from 0 to approximately \SI{18.1}{m.s^{-1}}.}
The corresponding sampled covariance for the normalized slow-time grid $\mathbf{t}$ follows the same CGP covariance model defined earlier:
\begin{align}\label{eq:ambiguity_covariance}
    \mathbf{K}_{\theta,\mathbf{t}}
    & =
    P\Big[
    \exp(-2\uppi^2\sigma_{fn}^2(t_a-t_b)^2) \\ \nonumber
    & \times \exp(j2\uppi\mu_{fn}(t_a-t_b))
    \Big]_{a,b}
    +\sigma_n^2\mathbf{I}.
\end{align}
Thus, the sampling sequence enters through all pairwise lags $t_a-t_b$, not only through the adjacent intervals. A candidate spectrum $\theta$ is ambiguous with a true spectrum $\theta_0$ when the two covariance matrices are statistically difficult to distinguish.

We measure this distinguishability with the Kullback-Leibler (KL) divergence \cite{Kullback1951,CoverThomas2006}. For two probability models, the KL divergence is the expected log ratio between the true density and the candidate density,
\begin{equation}
    D_{\mathrm{KL}}(p_{\theta_0}\|p_\theta)
    =
    \mathbb{E}_{\theta_0}
    \left[
    \log\frac{p_{\theta_0}(\mathbf{z})}{p_\theta(\mathbf{z})}
    \right].
    \label{eq:kl_definition}
\end{equation}
Intuitively, it is the average log-likelihood penalty incurred when data generated by the true model $\theta_0$ are explained by the candidate model $\theta$.
In the present case, both densities are zero-mean circular complex Gaussian densities,
\begin{equation}
    p_\theta(\mathbf{z})
    =
    \frac{1}{\uppi^N|\mathbf{K}_{\theta,\mathbf{t}}|}
    \exp\!\left(
    -\mathbf{z}^{H}
    \mathbf{K}_{\theta,\mathbf{t}}^{-1}
    \mathbf{z}
    \right).
    \label{eq:complex_gaussian_density}
\end{equation}
Substituting \eqref{eq:complex_gaussian_density} into \eqref{eq:kl_definition} gives a log-determinant term and a quadratic-form term. Taking the expectation under the true model uses, for any fixed $N\times N$ matrix $\mathbf{A}$ that does not depend on the random vector $\mathbf{z}$,
$\mathbb{E}_{\theta_0}\{\mathbf{z}^H\mathbf{A}\mathbf{z}\}
=\mathrm{tr}(\mathbf{A}\mathbf{K}_{\theta_0,\mathbf{t}})$, which yields the complex Gaussian KL divergence
\begin{equation}
    D_{\mathrm{KL}}(\theta_0\|\theta;\mathbf{t})
    =
    \log\frac{|\mathbf{K}_{\theta,\mathbf{t}}|}
    {|\mathbf{K}_{\theta_0,\mathbf{t}}|}
    +
    \mathrm{tr}\!\left(
    \mathbf{K}_{\theta,\mathbf{t}}^{-1}
    \mathbf{K}_{\theta_0,\mathbf{t}}
    \right)
    -N .
    \label{eq:cgp_kl}
\end{equation}

The KL divergence itself is dimensionless and is not a dB-valued quantity. However, it directly determines the expected relative likelihood. From \eqref{eq:kl_definition},
\begin{equation}
    \mathbb{E}_{\theta_0}\{\log p_\theta(\mathbf{z})\}
    =
    \mathbb{E}_{\theta_0}\{\log p_{\theta_0}(\mathbf{z})\}
    -
    D_{\mathrm{KL}}(\theta_0\|\theta;\mathbf{t}).
    \label{eq:expected_likelihood_kl}
\end{equation}
Thus, $D_{\mathrm{KL}}/N$ is the expected per-sample log-likelihood loss, which is dimensionless, caused by using the candidate spectrum. We display this loss as an average relative likelihood in decibels,
\begin{align}\label{eq:cgp_ambiguity}
    \mathcal{A}_{\mathbf{t}}(\theta;\theta_0)
    &=
    10\log_{10}
    \left[
    \exp\left(
    -\frac{D_{\mathrm{KL}}(\theta_0\|\theta;\mathbf{t})}{N}
    \right)
    \right] \\ \nonumber
    &=
    -\frac{10}{N\ln 10}
    D_{\mathrm{KL}}(\theta_0\|\theta;\mathbf{t}),
\end{align}
which has a maximum of zero at $\theta=\theta_0$. \revone{The value $D_{\mathrm{KL}}=0$, or equivalently $\mathcal{A}_{\mathbf{t}}=0$ dB, means that the candidate and true moment pairs induce indistinguishable sampled distributions. A positive KL divergence makes the candidate easier to reject and appears as a negative ambiguity level. The depth of the strongest remote sidelobe therefore gives a direct measure of protection against a wrong mean-width solution, while the main-lobe curvature describes local moment resolution.} This definition has two useful interpretations. First, \eqref{eq:cgp_ambiguity} is the average likelihood landscape before a particular noise realization is drawn. Second, for small perturbations,
\begin{equation}
    D_{\mathrm{KL}}(\theta_0\|\theta_0+\Delta\theta;\mathbf{t})
    =
    \frac{1}{2}\Delta\theta^T
    \mathbf{I}(\theta_0;\mathbf{t})
    \Delta\theta
    +o(\|\Delta\theta\|^2),
    \label{eq:kl_fisher}
\end{equation}
where $\mathbf{I}$ is the Fisher information matrix. Therefore, the local curvature of the same ambiguity surface controls the theoretical estimator variance, while remote high-valued sidelobes indicate global false basins that can cause branch errors and width bias. \revone{The proposed function thus joins local moment resolution and global branch ambiguity in one representation.}

\begin{figure*}[!t]
    \centering
    \paperfig[width=0.92\textwidth]{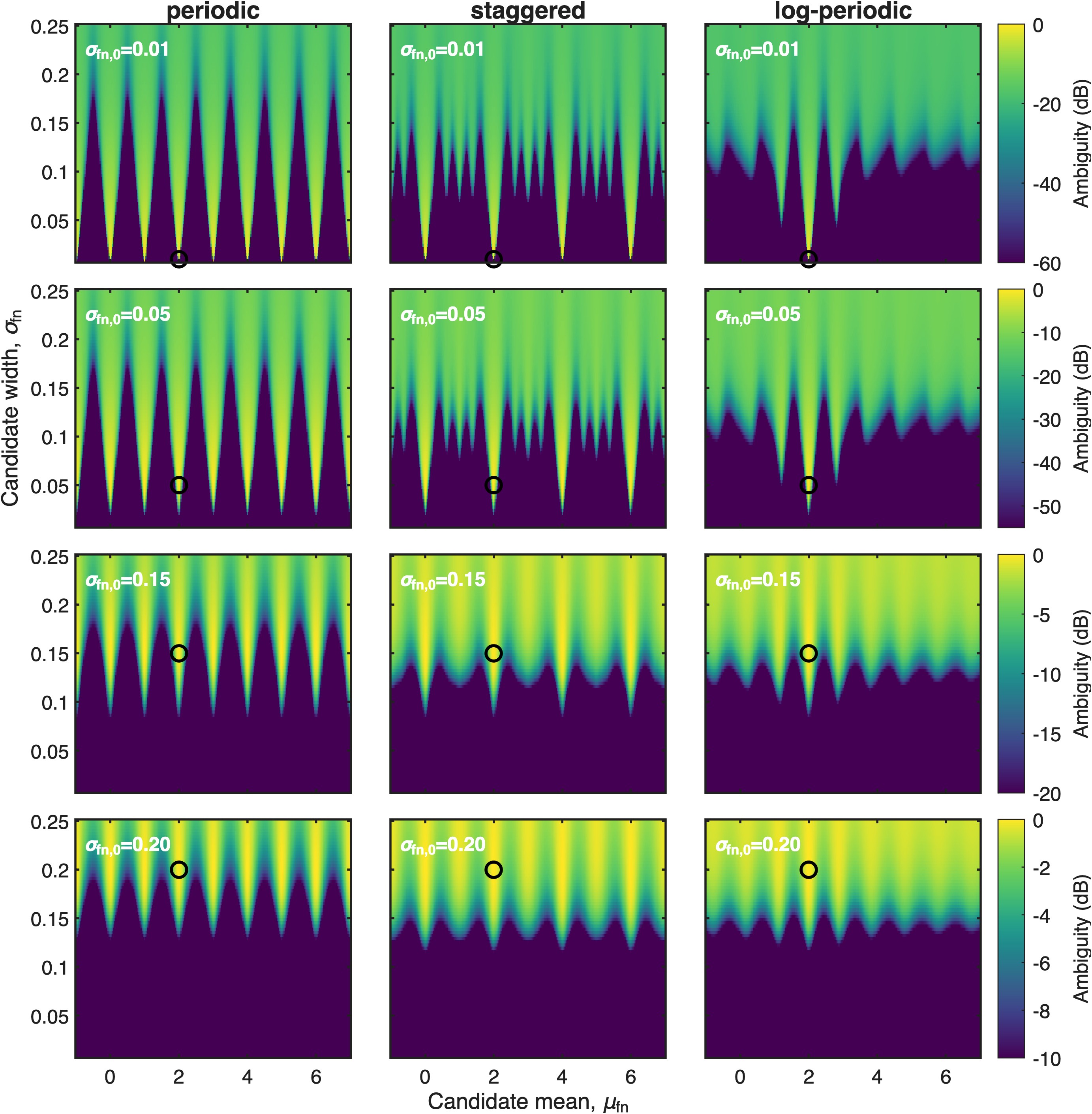}
    \caption{\revtwo{Moment-space ambiguity functions for a true mean Doppler $\mu_{fn,0}=2$ and true normalized widths $\sigma_{fn,0}\in\{0.01,0.05,0.15,0.20\}$. The columns show $\mathcal{A}_{\mathbf{t}}(\theta;\theta_0)$ over candidate mean and width for periodic, staggered, and log-periodic sampling. Each row uses its displayed ambiguity range so that both narrow- and broad-spectrum structure remains visible. The remote mean-width sidelobes are most strongly suppressed for narrow and moderate log-periodic spectra; broad spectra attenuate the informative long lags and reduce the sidelobe margin.}}
    \label{fig:theory_cgp_ambiguity}
\end{figure*}

\begin{figure*}[!t]
    \centering
    \paperfig[width=0.90\textwidth]{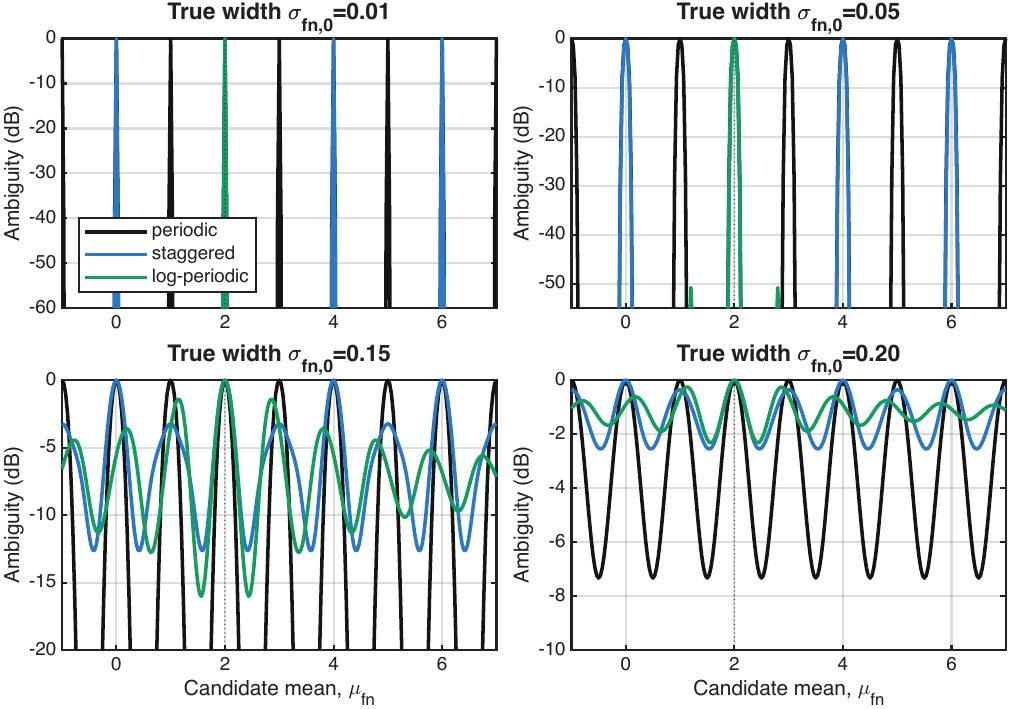}
    \caption{\revtwo{Mean-Doppler cuts through the moment-space ambiguity functions in Fig.~\ref{fig:theory_cgp_ambiguity} at the corresponding true spectral widths. The interval $\mu_{fn}\in[-1,7]$ exposes the competing branches around the true mean $\mu_{fn,0}=2$. Periodic sampling repeats the main lobe exactly; log-periodic sampling lowers the remote sidelobes for narrow and moderate spectra, while the three responses become less distinguishable as spectral width increases.}}
    \label{fig:theory_cgp_ambiguity_cuts}
\end{figure*}

Figs.~\ref{fig:theory_cgp_ambiguity} and \ref{fig:theory_cgp_ambiguity_cuts} show $\mathcal{A}_{\mathbf{t}}$ for a true mean $\mu_{fn,0}=2$, which lies outside the periodic Nyquist interval, and for four cases of true normalized spectral width. \revtwo{The two-dimensional maps cover the candidate moment domain $(\mu_{fn},\sigma_{fn})$, while the separate panels in Fig.~\ref{fig:theory_cgp_ambiguity_cuts} show readable cuts at each true width.} These panels do not show a spectral estimate from data; they show the ambiguity function implied by the sampling grid and the CGP covariance model. For narrow spectra, the periodic sequence produces exact repeated mean-Doppler branches. The staggered sequence reduces the simple integer repetition but retains strong structured sidelobes because many pairwise lags are generated from the same two interval lengths. The log-periodic sequence produces a more isolated main lobe around the true extended-Doppler spectrum. In the true-width cuts, its remote sidelobes remain below the main lobe at $\mu_{fn,0}$, so maximum-likelihood selection can reject competing branches. As the true spectral width increases, the Gaussian factor in \eqref{eq:ambiguity_covariance} suppresses long-lag covariance terms. The effective aperture in slow time is then reduced, the log-periodic sidelobe margin decreases, and all sampling strategies become less discriminating. This explains why extended-domain retrieval is easiest for narrow precipitation spectra and becomes intrinsically harder for broader weather spectra. \revboth{This width dependence is central to counter-aliasing. Our earlier spectral-estimation and counter-aliasing work showed that a finite dwell limits width estimation and that increasing normalized width weakens the covariance needed to distinguish mean-frequency branches \cite{Dash2024DopplerModel,Dash2024Counter-AliasingTrain}. The moment-space ambiguity function makes the two effects visible in one surface. In the limiting broad-spectrum case, the off-diagonal covariance terms in \eqref{eq:ambiguity_covariance} vanish (approaching the covariance of zero mean Gaussian white noise), and the covariance becomes independent of $\mu_{fn}$; no timing law can then recover the mean-frequency branch.}

\begin{figure*}[!t]
    \centering
    \paperfig[width=0.90\textwidth]{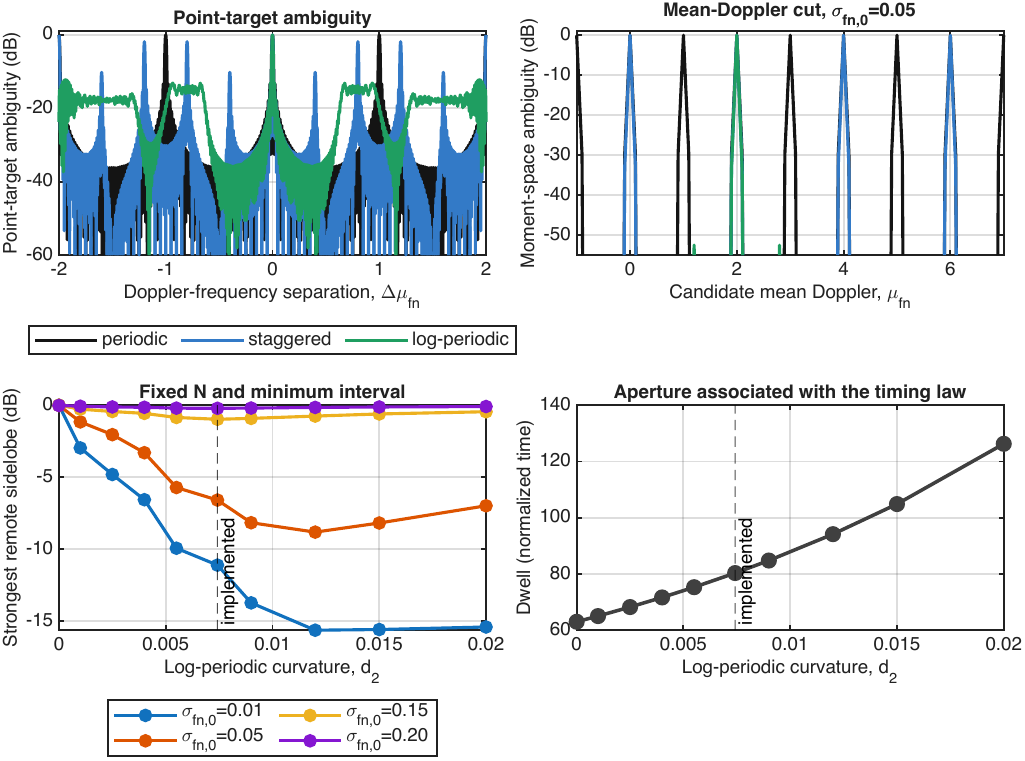}
    \caption{\revtwo{Point-target ambiguity and a moment-space mean-Doppler cut for the measured periodic, staggered, and log-periodic timing laws.} \revone{The lower panels show the strongest remote moment-space sidelobe as a function of the log-periodic curvature $d_2$ and the associated normalized dwell for $N=64$ with a fixed minimum interval. The dashed line marks the implemented value $d_2=0.007423$.}}
    \label{fig:ambiguity_design}
\end{figure*}

\revtwo{Fig.~\ref{fig:ambiguity_design} first compares the conventional point-target ambiguity responses of the three measured timing laws with a mean-Doppler cut through the proposed moment-space function. The point-target response describes the timing mask alone, whereas the moment-space response includes the loss of informative covariance at increasing spectral width.} \revone{The same figure shows how the log-periodic curvature $d_2$ controls the strongest remote mean-width sidelobe when $N$ and the minimum interval are fixed. For $\sigma_{fn,0}=0.01$, the sidelobe decreases from approximately 0~dB in the periodic limit to $-11.1$~dB at the implemented $d_2=0.007423$; for $\sigma_{fn,0}=0.05$, it decreases to $-6.6$~dB. The gain is smaller for broad spectra because their long-lag covariance is already weak. Increasing $d_2$ also lengthens the dwell, as shown in the fourth panel. The timing law is therefore tunable: $d_2$ can be selected according to the expected width range and the permissible dwell rather than treated as a fixed radar-specific constant.}

To connect the deterministic ambiguity surface with finite-data estimation, we also define a numerical false detection rate. For a fixed true parameter $\theta_0$ and sampling sequence $\mathbf{t}$, independent realizations $\mathbf{z}^{(r)}\sim\mathcal{CN}(0,\mathbf{K}_{\theta_0,\mathbf{t}})$ are generated. For each realization, the grid maximum-likelihood estimate is
\begin{equation}
    \hat{\theta}^{(r)}
    =
    \argmin_{\theta\in\Theta}
    \left[
    \log|\mathbf{K}_{\theta,\mathbf{t}}|
    +
    (\mathbf{z}^{(r)})^H
    \mathbf{K}_{\theta,\mathbf{t}}^{-1}
    \mathbf{z}^{(r)}
    \right].
    \label{eq:grid_mle_false_detection}
\end{equation}
The estimate is counted as a false detection if it falls outside a small neighborhood of the true spectrum,
\begin{equation}
    |\hat{\mu}_{fn}^{(r)}-\mu_{0, fn}|>\Delta_{\mu_{fn}}
    \quad\mathrm{or}\quad
    |\log(\hat{\sigma}_{fn}^{(r)}/\sigma_{0, fn})|>\Delta_{\sigma_{fn}} ,
    \label{eq:false_detection_rule}
\end{equation}
with $\Delta_{\mu_{fn}}=0.25$ and $\Delta_{\sigma_{fn}}=0.35$. The false detection rate is the fraction of Monte Carlo trials satisfying \eqref{eq:false_detection_rule}. This metric is deliberately stricter than a wrapped velocity error: it asks whether the estimator selected the correct extended-domain spectrum.

\revtwo{The candidate set is $\Theta=[-8,8]\times[0.008,0.35]$. The mean grid contains 161 uniformly spaced points with spacing 0.1, and the width grid contains 45 logarithmically spaced points. Each true parameter pair is evaluated using 200 independent realizations. Binomial 95\% confidence intervals are reported with the numerical results; they indicate the uncertainty caused by estimating a detection probability from a finite number of realizations.}

\begin{figure*}[!t]
    \centering
    \paperfig[width=0.92\textwidth]{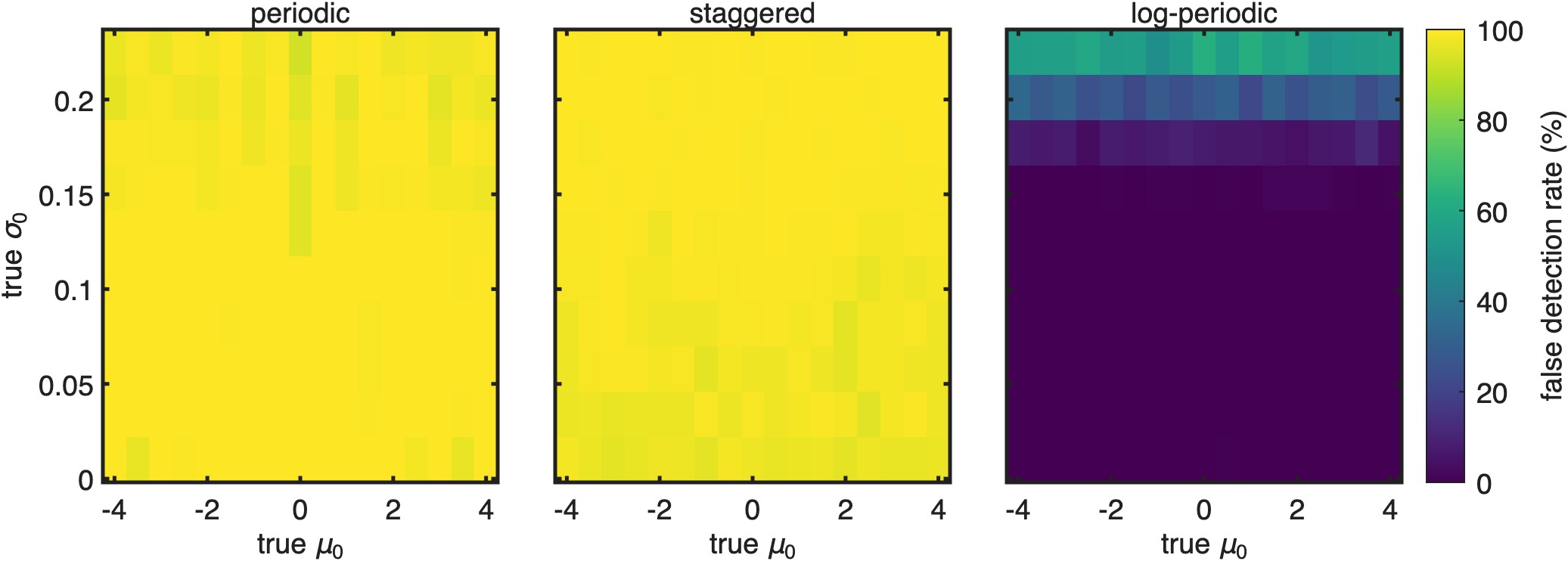}
    \caption{False detection rate from numerical CGP maximum-likelihood simulations over true mean Doppler and true spectral width. For each true parameter pair, 200 complex Gaussian slow-time realizations are generated and the maximum-likelihood estimate is found on the extended candidate grid. A trial is counted as false when the selected mean or width lies outside the neighborhood in \eqref{eq:false_detection_rule}. The log-periodic sequence has substantially lower false detection rate than the periodic and staggered sequences for narrow and moderate widths.}
    \label{fig:theory_false_detection}
\end{figure*}

Fig.~\ref{fig:theory_false_detection} shows the false detection rate over true mean and true spectral width for the three sampling sequences. The periodic case fails over most of the extended domain because its covariance is exactly periodic in $\mu_{fn}$ on the normalized unit grid. \revtwo{With $n_f=16$ indistinguishable periodic branches, unbiased tie breaking gives the analytical lower bound $1-1/n_f=93.75\%$ for selection outside the true branch.} The staggered case also has high false detection rates because its two-interval structure leaves strong extended-domain basins. In contrast, the log-periodic sequence has low false detection rates for narrow and moderate weather widths over the tested mean-Doppler range. In the 200-realization sweep, its maximum false detection rate over the tested means is at most $1\%$ for $\sigma_{fn,0}\leq 0.15$. \revtwo{For two events in 200 realizations, the one-sided 95\% binomial upper confidence limit is approximately 3\%, so the finite Monte Carlo count does not imply a zero-error operating region.} The mean rate increases to $7.1\%$ at $\sigma_{fn,0}=0.175$ and $27.7\%$ at $\sigma_{fn,0}=0.20$. \revboth{The rapid increase begins near $\sigma_{fn,0}\simeq0.16\approx1/(2\uppi)$. This is a practical transition: beyond it, progressively fewer nonzero lags retain enough covariance to reject remote branches, and the sequence approaches the intrinsically ambiguous broad-spectrum limit. Across the full tested interval $\mu_{fn}\in[-8,8]$, the low false-detection rate through $\sigma_{fn,0}=0.15$ demonstrates mean-branch discrimination over 16 conventional periodic Nyquist intervals for a distributed weather spectrum. To our knowledge, this joint mean-width ambiguity result has not previously been demonstrated in weather radar.} The hidden timing clock defines a much larger formal repetition interval, while spectral width and SNR determine how much of that interval is practically identifiable. These results show that the advantage of log-periodic counter-aliasing is not merely an increase in the formal clock-grid Nyquist interval; it is a reduction of the covariance-domain ambiguity between physically extended Doppler spectra.

\section{Estimators Compared}
\label{sec:estimators}
\subsection{Extended DFT and NUDFT}
\revtwo{For the periodic and log-periodic sequences, $\hat{S}_{\mathrm{F}}$ below denotes the common direct Fourier-periodogram expression. Its periodic-grid evaluation is the DFT, and its nonuniform log-periodic-grid evaluation is the NUDFT.} The spectrum is evaluated on the same extended grid $\mathcal{F}_{n_f}$:
\begin{equation}
    \revtwo{\hat{S}_{\mathrm{F}}(f)}
    =
    \frac{1}{N}
    \left|
    \sum_{k=0}^{N-1} z(t_k)\exp(-j2\uppi ft_k)
    \right|^2.
    \label{eq:nufft}
\end{equation}
For periodic sampling, this is the DFT evaluated on a dense extended axis. For log-periodic sampling, it is the Schuster periodogram for nonuniform time samples (NUDFT). The strongest peak location $\hat{f}_p$ is found on the extended grid, and the spectral moments are computed over one periodic Nyquist-width neighborhood,
\begin{equation}
    \mathcal{W}_{\hat{f}_p}
    =
    [\hat{f}_p-0.5,\hat{f}_p+0.5].
    \label{eq:dft_window}
\end{equation}
After subtracting the known noise floor, the mean and width are
\begin{align}
    \revtwo{\hat{\mu}_{fn, \mathrm{F}}}
    &=
    \frac{\int_{\mathcal{W}_{\hat{f}_p}} f \times 
    \left(\revtwo{\hat{S}_{\mathrm{F}}(f)}-\sigma_n^2\right)\,\mathrm{d}f}
    {\int_{\mathcal{W}_{\hat{f}_p}}
    \left(\revtwo{\hat{S}_{\mathrm{F}}(f)}-\sigma_n^2\right)\,\mathrm{d}f},\\
    \revtwo{\hat{\sigma}_{fn, \mathrm{F}}^2}
    &=
    \frac{\int_{\mathcal{W}_{\hat{f}_p}} (f-\revtwo{\hat{\mu}_{fn, \mathrm{F}}})^2
    \left(\revtwo{\hat{S}_{\mathrm{F}}(f)}-\sigma_n^2\right)\,\mathrm{d}f}
    {\int_{\mathcal{W}_{\hat{f}_p}}
    \left(\revtwo{\hat{S}_{\mathrm{F}}(f)}-\sigma_n^2\right)\,\mathrm{d}f}.
    \label{eq:dft_moments}
\end{align}
No wrapping is applied to these estimates in the results. The integrals here are approximated using summations because we have finite-frequency grids.
\revtwo{The calibrated noise level is subtracted from every periodogram bin without setting negative residual bins to zero. If the resulting second central moment is negative, only the final variance is clipped to zero before taking the square root. The Fourier estimates retain the convolution with the finite observation window; consequently, their widths are interpreted as finite-window periodogram moments.}

\subsection{Staggered PRT Velocity-Difference Estimator}
For the staggered sequence, two lag-one autocorrelation estimates are formed from the two alternating intervals. Let
\begin{equation}
    \hat{f}_1=\frac{1}{2\uppi \Delta t_1}\angle \hat{R}_1,
    \qquad
    \hat{f}_2=\frac{1}{2\uppi \Delta t_2}\angle \hat{R}_2,
    \label{eq:staggered_pair}
\end{equation}
where $\hat{R}_1$ and $\hat{R}_2$ are the corresponding pulse-pair correlations. The branch sequence of the velocity-difference method is generated from the constituent unambiguous frequencies $1/(2\Delta t_1)$ and $1/(2\Delta t_2)$. The selected branch minimizes
\begin{equation}
    \ell^\star =
    \argmin_{\ell} \left|
    (\hat{f}_1-\hat{f}_2)-C_\ell
    \right|,
    \label{eq:staggered_branch}
\end{equation}
Here $C_\ell$ is the expected difference between the two wrapped pulse-pair estimates for ambiguity branch $\ell$. Let $(p_\ell,q_\ell)\in\mathbb{Z}^2$ be the integer ambiguity corrections associated with the two constituent PRTs. Then
\begin{equation}
    C_\ell
    =
    \frac{q_\ell}{\Delta t_2}
    -
    \frac{p_\ell}{\Delta t_1},
    \qquad
    \hat{f}_{\mathrm{stag},\ell}
    =
    \hat{f}_2+\frac{q_\ell}{\Delta t_2}.
    \label{eq:staggered_branch_sequence}
\end{equation}
The implemented sequence starts from $(p_0,q_0)=(0,0)$ and alternately increments $q_\ell$ and $p_\ell$, together with the corresponding negative branches, until the full transfer-function interval is covered. After selecting $\ell^\star$ in \eqref{eq:staggered_branch}, the dealiased estimate is $\hat{f}_{\mathrm{stag},\ell^\star}$. \revtwo{This is the standard velocity-difference estimator for staggered PRT data, evaluated here for the high-precision measured ratio in \eqref{eq:staggered_ratio}.}

\subsection{CGP Multi-Start Maximum-Likelihood Estimation}
The reference CGP maximum-likelihood estimate (MLE) minimizes \eqref{eq:nll} over the extended domain
\begin{equation}
    \hat{\theta}_{\mathrm{CGP}}
    =
    \argmin_{\mu_{fn}\in[-n_f/2,n_f/2],\,
    \sigma_{fn}\in[\sigma_{\min},\sigma_{\max}]}
    \mathcal{L}(\mu_{fn},\sigma_{fn}),
    \label{eq:cgp_mle}
\end{equation}
with $\sigma_{\min}=0.005$ and $\sigma_{\max}=0.5$. In the full CGP reference, the constrained optimization is repeated from many random starting points, and the solution with the smallest negative log likelihood is retained. This reference is accurate but not ideal for real-time use, because random multi-start search makes the computational load variable.

\subsection{Fast CGP Basin Selection}
The fast CGP inversion uses the same likelihood and the same final constraints as \eqref{eq:cgp_mle}; only the initialization is changed. Let $\mathcal{P}$ be the set of the strongest NUDFT peak locations on the extended grid. A deterministic candidate set is formed as
\begin{equation}
    \mathcal{C}_0 =
    \left\{
    (p+m+\delta,s):
    p\in\mathcal{P},\;
    m\in\mathcal{M},\;
    \delta\in\mathcal{D},\;
    s\in\mathcal{S}_{\sigma}
    \right\}
    \label{eq:candidate_set}
\end{equation}
where each ordered pair is a candidate $(\mu_{fn},\sigma_{fn})$. The quantity $p\in\mathcal{P}$ is one of the dominant NUDFT peak locations; $m\in\mathcal{M}\subset\mathbb{Z}$ is an integer periodic-Nyquist translation; $\delta\in\mathcal{D}$ is a small local mean-Doppler offset around the translated peak; and $s\in\mathcal{S}_{\sigma}$ is a trial normalized spectral width. The symbol $s$ is used in \eqref{eq:candidate_set} to distinguish a discrete trial width from the unknown continuous parameter $\sigma_{fn}$ optimized at the final stage. \revtwo{The ten strongest NUDFT peaks are retained with a minimum normalized separation of 0.15.} In the implementation, $\mathcal{M}=\{-3,\ldots,3\}$, $\mathcal{D}=\{-0.08,-0.04,0,0.04,0.08\}$, and $\mathcal{S}_{\sigma}=\{0.008,0.012,0.016,0.024,0.04,0.07,0.11\}$. Only candidates with $\mu_{fn}\in\mathcal{F}_{n_f}$ are kept. Thus, the initializer uses only peaks observed in the current dwell, their ambiguity translates, local mean-Doppler offsets, and a small set of spectral-width trials.

\begin{figure*}[!t]
    \centering
    \paperfig[width=0.90\textwidth]{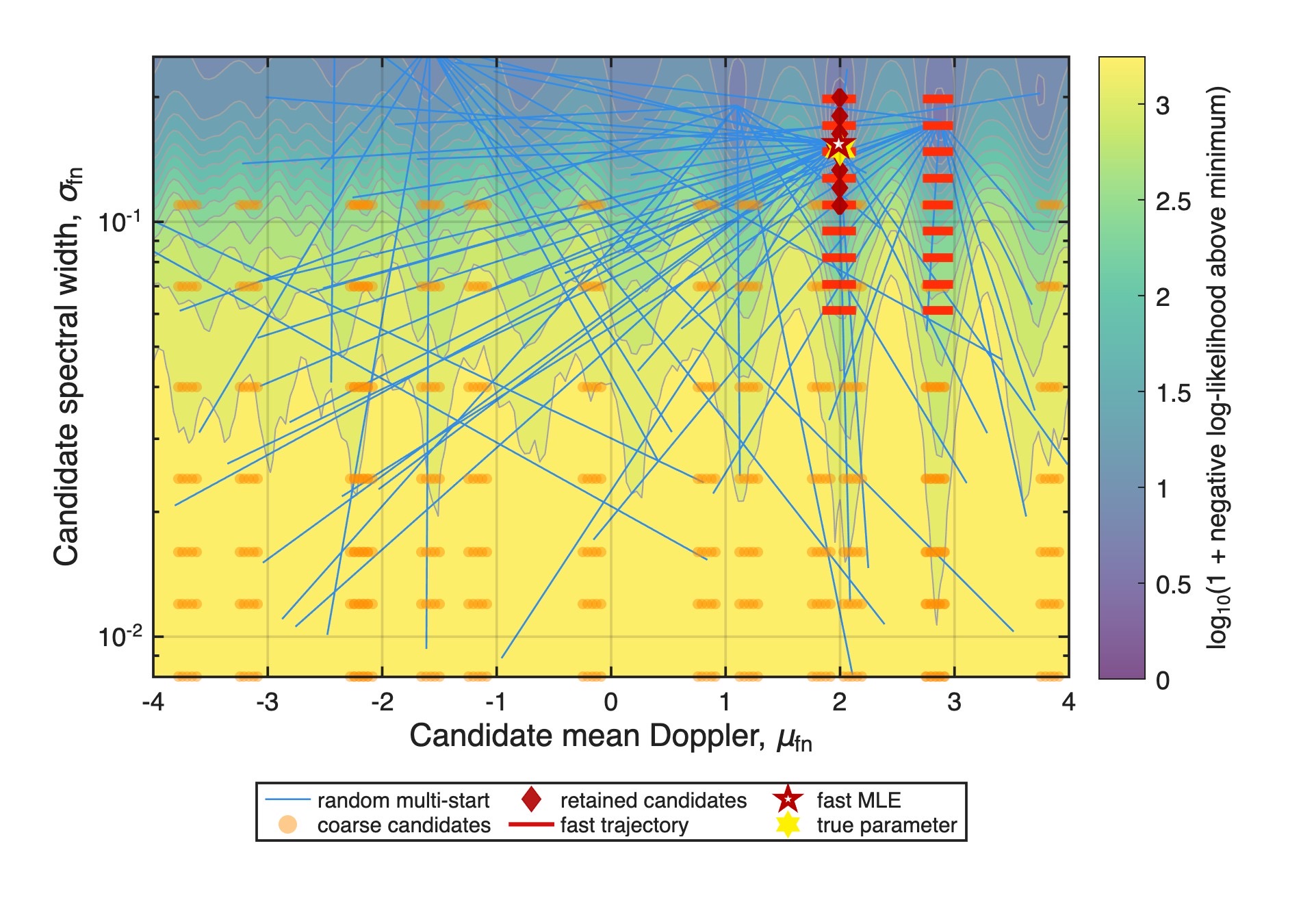}
    \caption{CGP initialization strategies shown on the same log-periodic likelihood surface for a synthetic target with $\mu_{fn,0}=2$ and $\sigma_{fn,0}=0.15$. The contours show $\log_{10}(1+\mathcal{L}-\mathcal{L}_{\min})$ over candidate mean Doppler and spectral width. Blue trajectories show conventional random multi-start optimization. Red and orange markers show the deterministic candidate set, retained refinements, and the single final fast-CGP trajectory. The true parameter is marked separately.}
    \label{fig:fast_cgp_initialization}
\end{figure*}

The candidates are ranked by \eqref{eq:nll}. The ten best candidates ($K_1=10$) are refined on a first local grid, and the three best resulting basins ($K_2=3$) are refined on a second narrower grid. \revtwo{The first grid spans $\mu_{fn}\pm0.035$ with spacing 0.005 and nine logarithmically spaced widths between the candidate width divided and multiplied by 1.8. The second grid spans $\mu_{fn}\pm0.012$ with spacing 0.0025 and seven logarithmically spaced widths within a factor 1.35. Widths are clipped to $[0.005,0.5]$.} The best refined point initializes one constrained \texttt{patternsearch} optimization of \eqref{eq:nll}. The implementation uses the \texttt{patternsearch} routine from the MathWorks Optimization Toolbox \cite{Mathworks2024OptimizationR2024a}, with the bounds in \eqref{eq:cgp_mle}, suppressed display, and the remaining options at their toolbox defaults. In brief, this derivative-free direct-search optimizer polls trial points around the current iterate and refines the search mesh when no lower objective value is found, which makes it suitable for this low-dimensional likelihood surface without requiring gradients. Fig.~\ref{fig:fast_cgp_initialization} shows this process on a representative two-parameter log-periodic CGP likelihood. The deterministic initial set samples several basins in both mean and spectral width, whereas the retained and refined sets concentrate around the likelihood-supported branch before the final optimizer is launched. The same likelihood panel also shows the conventional 64-start reference through its random starts, optimization trajectories, and converged endpoints. Algorithm \ref{alg:fast_cgp} summarizes the deterministic procedure.

\begin{algorithm}
\caption{Fast CGP inversion}
\label{alg:fast_cgp}
\begin{algorithmic}[1]
\State Evaluate $\hat{S}_{\mathrm{NUDFT}}(f)$ on $\mathcal{F}_{n_f}$.
\State Form $\mathcal{C}_0$ from NUDFT peaks, ambiguity translates, local offsets, and trial widths.
\State Score every candidate using the raw CGP negative log likelihood in \eqref{eq:nll}.
\State Keep the best $K_1$ candidates and build a first local grid around them.
\State Score the first local grid and keep the best $K_2$ candidates.
\State Build and score a second, narrower local grid.
\State Start one constrained optimizer from the best refined candidate.
\State Return the final raw maximum-likelihood solution.
\end{algorithmic}
\end{algorithm}

\section{Frequency-Domain Spectrum Reconstruction}
\label{sec:spectrum_reconstruction}
After estimating $\hat{\theta}$, the CGP posterior is evaluated directly on a frequency grid. The prior spectral density corresponding to \eqref{eq:single_covariance} is given in \eqref{eq:ambiguity_spectrum}.
Following the usual Gaussian conditioning form used for GP spectral reconstruction \cite{Dash2023PrecipitationPrior,Tobar2018BayesianEstimation}, the posterior mean at the frequency grid $\mathbf{f}$ can be written as
\begin{equation}
    \bar{\mathbf{s}}_{\mathbf{f}|\mathbf{z}}
    =
    \mathbf{K}_{\mathbf{f}\mathbf{t}}
    \mathbf{K}_{\mathbf{t}\mathbf{t}}^{-1}
    \mathbf{z},
    \label{eq:posterior_mean}
\end{equation}
where $\mathbf{K}_{\mathbf{t}\mathbf{t}}=\mathbf{K}_{\hat{\theta}}$. \revtwo{For frequency $f_m$ and time $t_k$, the cross-covariance used here is
\begin{equation}
 [\mathbf{K}_{\mathbf{f}\mathbf{t}}]_{m,k}
 =S_{\hat{\theta}}(f_m)\exp(j2\uppi f_m t_k),
 \label{eq:frequency_time_cross_covariance}
\end{equation}
with the conjugate convention implied by \eqref{eq:single_covariance}} \cite[Eq.~(19)]{Dash2023PrecipitationPrior}, \cite[Eq.~(13-14)]{Dash2024Counter-AliasingTrain}. \revtwo{Equation~\eqref{eq:posterior_mean} is the posterior mean conditioned on the fitted parameter vector $\hat{\theta}$. Its concentration and width therefore reflect the fitted Gaussian weather-spectrum model.} This reconstruction is performed directly in the Doppler domain. No slow-time interpolation is used, and the same Cholesky factorization used for the likelihood can be reused for posterior evaluation. The size of the covariance matrix being inverted in this case depends only on the number of samples currently present in the dwell, which can be very small.

\section{Real-Data Analysis and Preprocessing}
\label{sec:real_data}
The measurements were collected on 8 June 2026 with the Polarimetric Agile Radar in S- and X-band (PARSAX) S-band FMCW radar operated by the Microwave Sensing, Signals and Systems (MS3) group at Delft University of Technology \revtwo{\cite{Krasnov2010PARSAX}}. The radar was pointed vertically during the considered precipitation event. Hence, the Doppler spectra are dominated by hydrometeor fall velocity and vertical air motion, and no ground clutter component is expected in the processed range cells. PARSAX is a polarimetric radar, but for this Doppler counter-aliasing experiment only the horizontal-transmit horizontal-receive (HH) channel was stored and processed. The carrier frequency is \SI{3.315}{GHz}. The first slow-time interval was \SI{0.5}{ms}, corresponding to a periodic normalized Nyquist interval of $[-0.5,0.5]$.

Three acquisitions are used: periodic, staggered, and log-periodic slow-time sampling. The periodic, log-periodic, and staggered acquisitions used in the main comparison started at 16:02:51, 16:04:38, and 16:05:56 Coordinated Universal Time (UTC), respectively. They were not simultaneous and therefore contain a small time separation. The precipitation structure in height remained similar during the measurement interval, allowing the three acquisitions to be compared as realizations of the same weather situation. \revboth{Because atmospheric evolution cannot be separated completely from timing-law differences in these sequential measurements, this comparison is used to establish practical operation and central-branch stability rather than a controlled acquisition-only performance difference. The event did not contain precipitation velocities outside the periodic Nyquist interval. Controlled branch recovery is evaluated separately by phase modulation of each measured record.}

For each chirp, the fast-time data are mean-subtracted, Hamming-windowed, and transformed with an 8192-point range FFT. The positive range cells are retained, giving 4096 processed range bins. The measurements use a \SI{50}{MHz} chirp bandwidth, corresponding to a nominal range resolution of approximately \SI{3}{m}. The chirp length was $2^{13}\times 27+10$ samples at an arbitrary waveform generator (AWG) sampling rate of \SI{500}{MHz}, i.e., approximately \SI{442.4}{\micro\second}.

The slow-time timing headers are recorded on a separate verified \SI{120}{MHz} timing clock. The integer timestamp represents the number of clock cycles separating the first acquisition trigger from the trigger of the snapshot of interest, so one recorded timing tick is approximately \SI{8.33}{ns}. The novelty of the measurement design is that the aperiodic experiment changes the waiting/reset time between otherwise unchanged FMCW chirps; the chirp waveform, chirp bandwidth, fast-time sampling rate, and range FFT are kept fixed. \rev{Therefore, the experiment does not trade Doppler ambiguity reduction against a shorter chirp or a modified range-processing chain. It changes the slow-time schedule while preserving the fast-time waveform that provides the range information.}

Although the data contain 4096 positive range bins, the Doppler analysis is restricted to the range interval where precipitation is visually present in the measured power profiles. \revtwo{The positive range-FFT axis extends to approximately \SI{12.3}{km}, whereas cells 40-1200 used in the quantitative comparison span approximately \SI{0.12}{km} to \SI{3.60}{km}.} This interval is much smaller than the non-overlaid range implied by the minimum slow-time interval, $cT_0/2\approx\SI{75}{km}$; range-overlaid signals are outside the scope of the present experiment. Because only the waiting time is changed, the nominal range resolution and range-processing limits are identical for all three acquisitions.

\revtwo{The PSE is also applied to the measured periodic record. Its finite-window Whittle likelihood is evaluated on the principal normalized interval $[-0.5,0.5]$, using the calibrated noise variance and the average signal power after noise subtraction. This gives a directly applicable periodic-data reference for Doppler mean and spectral width. Because every integer-shifted mean produces the same periodically sampled data distribution, the PSE result is reported within the principal interval and is not assigned an extended-domain branch.}

\subsection{Controlled Doppler-Branch Injection}
\revboth{A known Doppler translation is applied to the measured complex slow-time echo in each range cell according to
\begin{equation}
 z_{f_s}(t_k)=z(t_k)\exp(j2\uppi f_s t_k),
 \label{eq:phase_injection}
\end{equation}
where $f_s$ is normalized by the same first slow-time interval as the estimator. This operation translates the complete measured weather spectrum by $f_s$ while retaining its instantaneous amplitudes, spectral fluctuations, SNR, and measured timing grid. Let $\mu_{fn,r}^{\star}(0)$ denote the unknown physical mean in retained range cell $r$ before injection. The imposed, and therefore exactly known, translated mean is
\begin{equation}
 \mu_{fn,r}^{\star}(f_s)=\mu_{fn,r}^{\star}(0)+f_s,
 \qquad
 \Delta\mu_{fn,r}^{\star}=f_s.
 \label{eq:injection_truth}
\end{equation}
The absolute mean $\mu_{fn,r}^{\star}(0)$ is not assumed known. Instead, for any estimator, $\hat{\mu}_{fn,r}^{(0)}$ and $\hat{\mu}_{fn,r}^{(s)}$ are its estimates from the original and translated records, respectively, and its recovered increment is
\begin{equation}
 \Delta\hat{\mu}_{fn,r}(f_s)=\hat{\mu}_{fn,r}^{(s)}-\hat{\mu}_{fn,r}^{(0)}.
 \label{eq:recovered_increment}
\end{equation}
Thus the known reference is $f_s$, not the output of a particular estimator. For each estimator and imposed shift, let $\mathcal{G}_s$ be the set of range cells that pass the SNR gate and have finite estimates before and after injection, and let $N_s=|\mathcal{G}_s|$ be its number of cells. Correct branch recovery is evaluated using
\begin{equation}
 \begin{aligned}
 \mathcal{B}_{r,s}&=\mathds{1}\!\left\{
 \begin{gathered}
 \operatorname{round}\!\left(\hat{\mu}_{fn,r}^{(s)}\right) =\operatorname{round}\!\left(\hat{\mu}_{fn,r}^{(0)}+f_s\right)
 \end{gathered}\right\},\\
 \mathcal{I}_{r,s}&=\mathds{1}\!\left\{
 \left|\Delta\hat{\mu}_{fn,r}(f_s)-f_s\right|\leq0.25\right\}\mathcal{B}_{r,s},\\
 R_s&=\frac{100}{N_s}\sum_{r\in\mathcal{G}_s}\mathcal{I}_{r,s},
 \end{aligned}
 \label{eq:branch_recovery_rate}
\end{equation}
where $\mathds{1}$ is the indicator function. Here $\mathcal{B}_{r,s}$ is the expected-branch indicator: it is one when the translated estimate occupies the integer branch predicted by the estimator's unshifted estimate plus the imposed shift, and zero otherwise. Thus, $\mathcal{I}_{r,s}$ requires both agreement of the recovered shift with $f_s$ and agreement with that expected branch. With the already defined full periodic Nyquist velocity span $2V_a^{(\mathrm{per})}=\SI{90.5}{m.s^{-1}}$ per normalized Doppler unit, the root-mean-square error (RMSE) of the recovered increment is
\begin{equation}
 \operatorname{RMSE}_{v}(f_s)=2V_a^{(\mathrm{per})}\sqrt{\frac{1}{N_s}\sum_{r\in\mathcal{G}_s}
 \left[\Delta\hat{\mu}_{fn,r}(f_s)-f_s\right]^2}.
 \label{eq:injection_rmse}
\end{equation}
It is therefore sensitive both to a common offset and to isolated wrong-branch selections.}

\revboth{The width change is also computed for each cell by subtracting its unshifted estimated width from its estimated width after translation. A value near zero means that the imposed phase modulation moved the retrieved spectrum without changing its estimated spread. The median of these cell-by-cell changes summarizes the typical behavior, while the 25th and 75th percentiles describe the middle half of the retained cells. The same shifts and precipitation-support gate are used for all three acquisitions.}

\revboth{The phase injection tests statistical identifiability on known extended-domain branches. It does not reproduce the range migration or FMCW range-Doppler displacement of a physically fast scatterer. The measured periodic, log-periodic, and staggered dwell durations are approximately \SI{31.5}{ms}, \SI{39.87}{ms}, and \SI{38.97}{ms}, respectively. Traversing one \SI{3}{m} range cell during these dwells corresponds to about \SI{95}{m.s^{-1}}, \SI{75}{m.s^{-1}}, and \SI{77}{m.s^{-1}}. For the \SI{50}{MHz}, \SI{442.4}{\micro\second} FMCW chirp, a Doppler-induced range displacement of one range cell occurs at approximately \SI{102}{m.s^{-1}}. At the edge of the displayed normalized domain, $|f_n|=8$ corresponds to approximately \SI{724}{m.s^{-1}}. Such a target would move through about 7.6-9.6 range cells during one measured dwell, depending on the timing law, and its FMCW Doppler displacement would be approximately \SI{21.2}{m}, or 7.1 range cells. The formal clock-grid search interval is therefore much wider than the interval over which a physical target can be treated as stationary within one range cell.}

\section{Results}
\label{sec:results}
\subsection{Known-Width Finite-Dwell Comparison}
\revtwo{The spectral-width comparison uses simulated Gaussian weather echoes with known moments. The measured periodic, log-periodic, and staggered timing grids are truncated to $N\in\{32,64\}$ samples. The true mean is $\mu_{fn}=0.15$, and 14 widths from $\sigma_{fn}=0.008$ to 0.25 are tested at SNRs of \SIlist{10;20;30}{dB}. Each point uses the same 500 random realizations for all estimators. The width $\sigma_{fn}$ is normalized by the full Nyquist span associated with the minimum slow-time interval $T_0$.}

\revtwo{The signal power is fixed at $P=1$, and the corresponding noise variance is known, as in \cite{Dash2024DopplerModel,Dash2024Counter-AliasingTrain}. PSE and CGP use both quantities. The DFT applied to periodic data, the NUDFT applied to log-periodic data, and the velocity-difference estimator applied to staggered data use the same known noise variance. The local CGP reference starts in the true mean-width basin and therefore measures the finite-record precision of the CGP likelihood. Fast CGP receives no such information and selects its starting basin from the observed log-periodic record.}

\revtwo{The periodic Fourier-bin spacing is approximately \SI{2.83}{m.s^{-1}} for $N=32$ and \SI{1.41}{m.s^{-1}} for $N=64$. The widths observed in the measurements, which are close to \SI{1}{m.s^{-1}}, are therefore below the direct Fourier resolution. For $\sigma_{fn}=0.011$, $1/\sigma_{fn}\approx91$ samples, so both record lengths represent the short-record case discussed in \cite{Dash2024DopplerModel}.}

\begin{figure*}[!t]
    \centering
    \paperfig[width=0.92\textwidth]{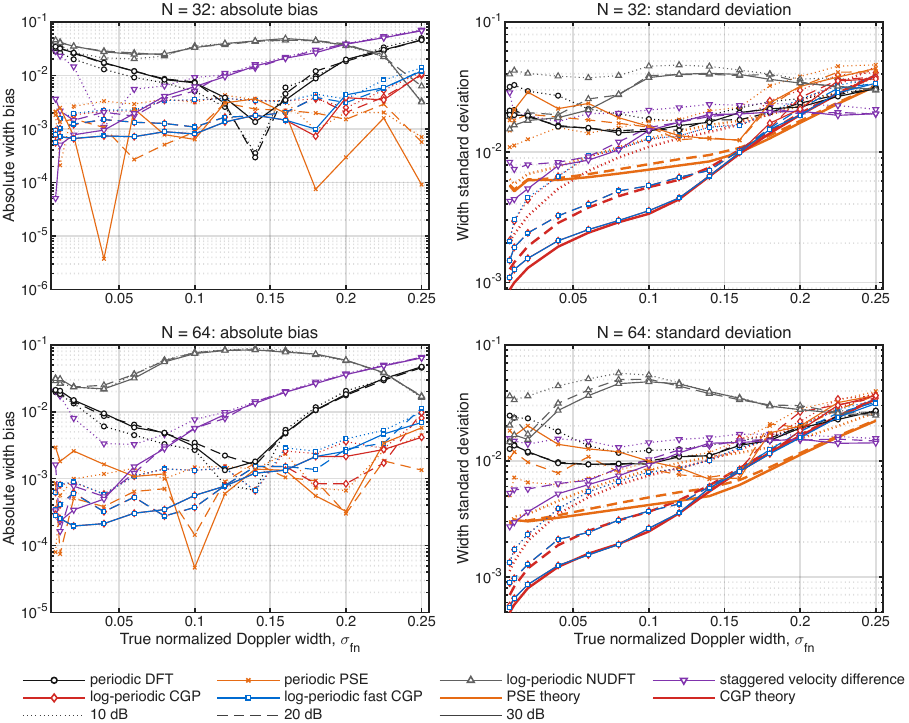}
    \caption{\revtwo{Known-width comparison in normalized Doppler units using 500 common random realizations for each true-width and SNR pair. Black circles denote DFT moments from periodic data, orange crosses denote PSE estimates from periodic data, gray upward triangles denote NUDFT moments from log-periodic data, purple downward triangles denote velocity-difference estimates from staggered data, red diamonds denote the local CGP reference on log-periodic data, and blue squares denote fast CGP on the same data. Dotted, dashed, and solid curves denote \SIlist{10;20;30}{dB}, respectively. The left panels show absolute width bias. The right panels show empirical standard deviation together with the marker-free periodic-PSE and log-periodic-CGP finite-record theoretical standard deviations.}}
    \label{fig:pse_width_benchmark}
\end{figure*}

\revtwo{Fig.~\ref{fig:pse_width_benchmark} shows the absolute width bias and standard deviation on logarithmic axes. The absolute bias, $|\mathbb{E}\{\hat{\sigma}_{fn}\}-\sigma_{fn}|$, measures the magnitude of systematic over- or underestimation. The standard deviation measures the variation among the 500 estimates. The marker-free curves show the theoretical width standard deviation $s_{\mathrm{th},\sigma}=\sqrt{[\mathbf{I}^{-1}]_{\sigma\sigma}}$, following Levin \cite[eq.~(4)]{Levin1965PowerEstimation} and the finite-record PSE derivation in \cite[Appendix~B]{Dash2024DopplerModel}. Here, $\mathbf{I}$ is the expected Fisher information matrix, which measures the local sensitivity of each likelihood to changes in the model parameters. The PSE curve follows from the approximate finite-window spectral likelihood on the periodic grid, whereas the CGP curve follows from the exact Gaussian likelihood of the log-periodic time samples. Each curve is evaluated separately for the stated record length, width, and SNR. Because the numerical estimators can be biased for finite $N$, these curves are interpreted as theoretical inverse sensitivities rather than attainable bounds for an unbiased estimator. The conventional unbiased bound is approached asymptotically as the bias vanishes \cite{Dash2024DopplerModel}.}

\revtwo{At \SI{20}{dB}, $\sigma_{fn}=0.011$ corresponds to a physical width of approximately \SI{1}{m.s^{-1}}. For $N=64$, the absolute biases of the periodic DFT and log-periodic NUDFT are 0.0184 and 0.0296, respectively. The biases of periodic PSE and the staggered velocity-difference estimator are below $2\times10^{-4}$, while the absolute bias of both log-periodic CGP implementations is $4.07\times10^{-4}$. The corresponding standard deviations are 0.0137, 0.0234, 0.0119, and 0.00565, respectively; both CGP implementations have a standard deviation of 0.000971.}

\revtwo{For $N=32$, the corresponding standard deviations are 0.0210 for the periodic DFT, 0.0201 for the log-periodic NUDFT, 0.0175 for periodic PSE, 0.00739 for the staggered velocity-difference estimator, and 0.00186 for both log-periodic CGP implementations. The shorter record increases the spread of every estimator, but fast CGP continues to follow the local CGP reference throughout the informative width range.}

\revtwo{These results separate the effects of the estimator and the sampling pattern. The large Fourier widths arise mainly from the finite observation window. Parametric PSE and CGP modeling can recover narrower widths. The role of log-periodic sampling is different: it reduces the competing ambiguity branches over the extended Doppler interval.}

\subsection{Spectral Reconstructions at Selected Range Cells}
\begin{figure*}[!t]
    \centering
    \paperfig[width=0.90\textwidth]{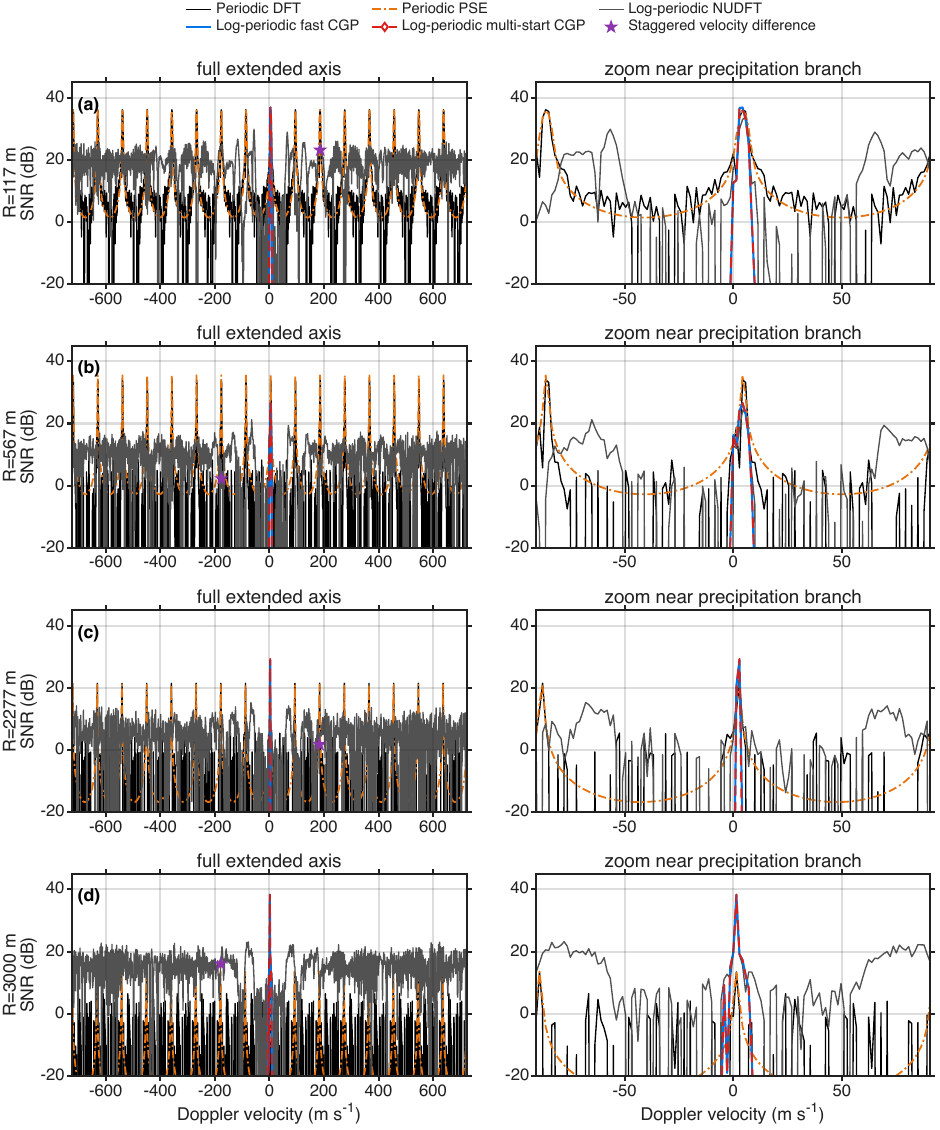}
    \caption{Extended Doppler velocity spectra for selected ranges. Each row combines the full extended velocity axis with a zoom around the precipitation Doppler region. \rev{Rows (a)-(d) correspond to the selected range cells from lowest to highest displayed range.} \revtwo{The periodic PSE spectrum is repeated across the extended axis because its ambiguity branch is unresolved.}}
    \label{fig:spectrum_cells_combined}
\end{figure*}

Fig.~\ref{fig:spectrum_cells_combined} compares the reconstructed spectra at representative ranges. Each row combines the full extended Doppler interval with a closer view of the precipitation region. The spectra show power above the calibrated noise level rather than absolute receiver power. \revtwo{The DFT spectrum from the periodic acquisition repeats at every periodic Nyquist interval. PSE gives a narrower fitted spectrum for the same periodic data, but this spectrum also repeats because periodic sampling cannot identify its branch. The NUDFT spectrum from the log-periodic acquisition removes the exact repetition but retains visible sidelobes. Both CGP implementations applied to the log-periodic data concentrate the power near the likelihood-supported branch and give similar spectra. The velocity-difference estimate from the staggered acquisition appears as a marker because this estimator returns moments rather than a complete spectrum.}

\subsection{Quantitative Moment and Branch Comparison}
\begin{figure*}[!t]
    \centering
    \paperfig[width=0.94\textwidth]{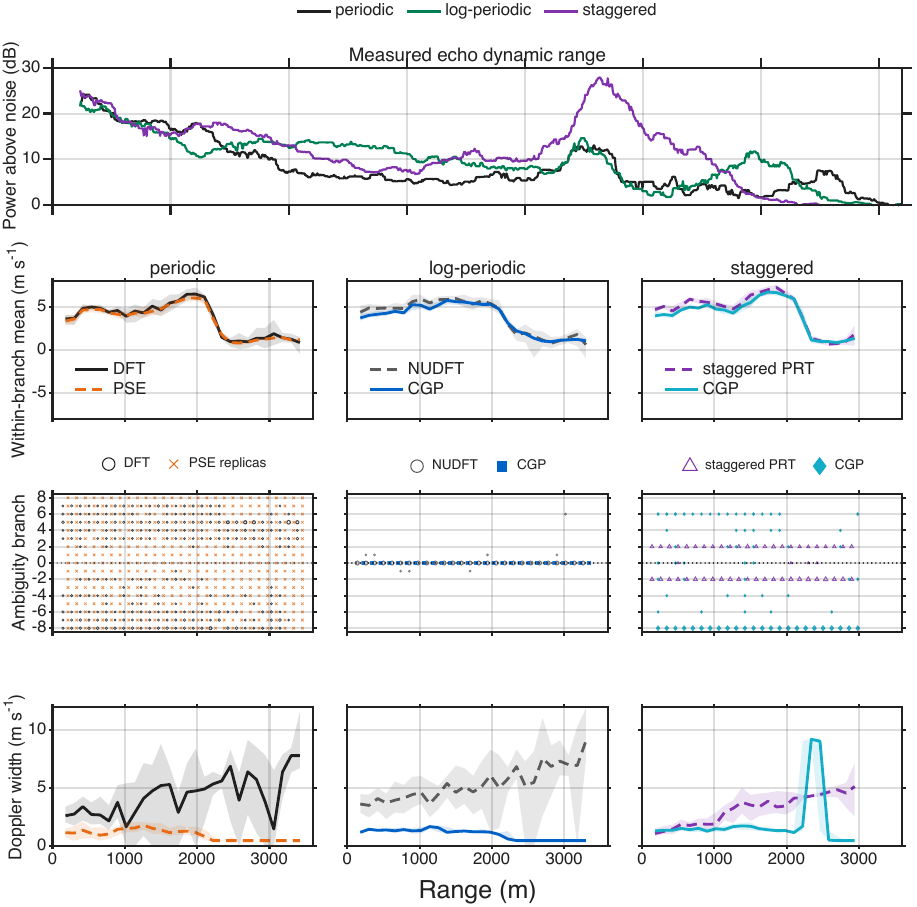}
    \caption{\revtwo{Range-resolved real-data comparison. The top panel shows power above the calibrated noise level. The second and fourth rows show the median within-branch mean velocity and Doppler width in \SI{120}{m} bins for cells with SNR $\geq\SI{3}{dB}$. Each shaded region covers the 25th-75th percentile range, containing the middle half of the retained cell estimates. The third row spans the extended branch domain $k=-8,\ldots,8$. For log-periodic and staggered sampling, marker area is proportional to the percentage of cells selecting a branch within each range bin. Paired estimators are displaced slightly to either side of the bin center for visibility, without changing their branch or physical range. In the log-periodic column, open and filled markers denote NUDFT and CGP selections. In the staggered column, they denote velocity-difference and CGP selections. In the periodic panel, DFT circles show numerical choices among equal maxima and orange crosses show the equivalent PSE replicas; neither identifies the periodic branch.}}
    \label{fig:real_data_profiles}
\end{figure*}

\revtwo{Fig.~\ref{fig:real_data_profiles} preserves the range dependence of the retrieved moments and their extended-domain branches. The top panel shows the measured echo strength. In the second and fourth rows, each line is the median within-branch mean velocity or Doppler width in a \SI{120}{m} range bin for cells with SNR $\geq\SI{3}{dB}$. The native 3-m range cells are grouped into non-overlapping 120-m intervals for visualization. Within each interval, the line gives the median of the SNR-gated estimates and the shaded region contains their 25th–75th percentiles; it is not a confidence interval for the median; it is not a confidence interval for the median. For methods that identify an extended-domain solution, the ambiguity branch is $k=\mathrm{round}(\hat{\mu}_{fn})$; $k=0$ denotes the principal precipitation branch and nonzero values denote remote branches. In the log-periodic and staggered columns, marker area in the third row is proportional to the percentage of retained cells selecting each branch in a range bin. In the periodic column, the DFT circles show numerical choices among exactly repeated maxima and the PSE crosses show all equivalent replicas; neither periodic display represents an identified branch.}

\begin{figure*}[!t]
    \centering
    \paperfig[width=0.88\textwidth]{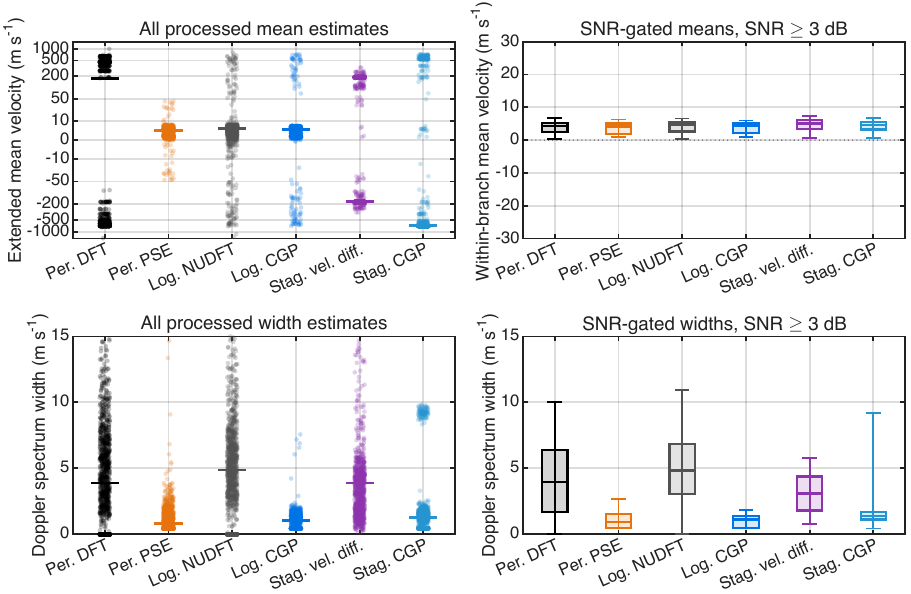}
    \caption{\revtwo{Distribution summary of the real-data Doppler moments. The left panels show all processed range cells; the right panels retain cells with SNR $\geq\SI{3}{dB}$. The upper-left panel keeps the extended velocity estimates visible, while the upper-right compares their positions within one Nyquist interval. The lower panels show Doppler widths. In the right panels, the line inside each box is the median, the box contains the middle half of the cells (25th-75th percentiles), and the whiskers span the 5th-95th percentiles. Periodic DFT selections on remote branches are choices among equal maxima rather than identified branches, and periodic PSE is confined to the principal interval.}}
    \label{fig:real_data_summary}
\end{figure*}

\revtwo{Fig.~\ref{fig:real_data_summary} gives a direct distribution view. The left panels retain all processed cells and show the extended mean estimates and Doppler widths. The periodic-DFT points on remote branches are numerical selections among equal periodic maxima, not identified branches; periodic PSE is evaluated only in the principal interval. The right panels retain cells whose echo power is at least \SI{3}{dB} above the calibrated noise level. For the mean comparison, $\hat{\mu}_{fn}=k+\delta$ is shifted by the integer branch $k$ into the principal interval, so the plotted velocity corresponds to $\delta\in[-0.5,0.5)$. This places the precipitation-scale velocities on a common interval without hiding remote selections in the all-cell panel.}

\revtwo{Table~\ref{tab:real_data_summary} gives the corresponding numerical values. For the 955 periodic cells, the median width decreases from \SI{3.96}{m.s^{-1}} with the DFT to \SI{0.94}{m.s^{-1}} with PSE. The middle half of the width estimates spans \SI{1.73}{m.s^{-1}} to \SI{6.38}{m.s^{-1}} for the DFT and \SI{0.45}{m.s^{-1}} to \SI{1.56}{m.s^{-1}} for PSE. Their median velocities within the principal interval are similar: \SI{4.27}{m.s^{-1}} for the DFT and \SI{4.16}{m.s^{-1}} for PSE. A branch percentage is not reported for either method because periodic sampling leaves all repeated branches equivalent.}

\revtwo{The log-periodic comparison gives a different result. The NUDFT selects the central branch for 99.1\% of the 979 retained cells, and the CGP selects it for all cells. Their median widths are \SI{4.85}{m.s^{-1}} and \SI{1.14}{m.s^{-1}}, respectively. Thus, the timing pattern supports branch selection, while the covariance model gives the narrower parametric width. Both estimators applied to the staggered data select several remote branches, despite the large formal interval implied by the recorded timestamps.}

\begin{table*}[t]
\centering
\caption{\revtwo{Real-data statistics for cells with SNR $\geq\SI{3}{dB}$. The periodic, log-periodic, and staggered acquisitions retain 955, 979, and 940 cells, respectively. Each bracket gives the 25th and 75th percentiles, i.e., the limits containing the middle half of the cell estimates. Mean velocity is reported on the extended domain when the branch is identifiable and in the principal interval for periodic DFT and PSE. A dash indicates that periodic sampling does not identify the ambiguity branch.}}
\label{tab:real_data_summary}
\begin{tabular}{lccc}
\toprule
Acquisition and estimator & \shortstack{Mean velocity, median\\(25th-75th percentile) (\si{m.s^{-1}})} & \shortstack{Width, median\\(25th-75th percentile) (\si{m.s^{-1}})} & \shortstack{Central branch\\$k=0$ (\%)} \\
\midrule
Periodic DFT       & 4.27 [2.55, 5.25]          & 3.96 [1.73, 6.38] & - \\
Periodic PSE       & 4.16 [1.99, 5.06]          & 0.94 [0.45, 1.56] & - \\
\revtwo{Log-periodic NUDFT} & 4.85 [2.68, 5.67]          & 4.85 [3.06, 6.87] & 99.1 \\
Log-periodic CGP   & 4.42 [2.29, 5.25]          & 1.14 [0.45, 1.41] & 100.0 \\
Staggered velocity difference & -174.80 [-176.59, 186.19]  & 3.11 [1.81, 4.39] & 0.4 \\
Staggered CGP      & -719.15 [-720.65, -717.85] & 1.37 [1.13, 1.67] & 1.2 \\
\bottomrule
\end{tabular}
\end{table*}

\revtwo{The analyzed height interval includes a pronounced power enhancement near \SI{2.2}{km} to \SI{2.5}{km}, accompanied by locally increased spectral width, particularly in the staggered acquisition. This behavior is consistent with a melting-layer signature.}

\revtwo{For the log-periodic acquisition, the zero-Doppler bin is a median \SI{15.4}{dB} below the strongest periodogram bin within the principal Nyquist interval for the SNR-gated cells; it lies within \SI{1}{dB} of the local maximum in only 1.7\% of those cells. The radar was pointed vertically, so the main beam sampled hydrometeors aloft rather than a stationary ground surface. The central-branch concentration is therefore not explained by a generally dominant static component.}

\subsection{Controlled Doppler-Branch Recovery}
\revboth{The normalized shifts $f_s\in\{0.5,1.5,2.5,3.5,4.5,5.5,6.5\}$ are applied to the measured complex echoes using \eqref{eq:phase_injection}. The same SNR gate as in Fig.~\ref{fig:real_data_summary} retains 955 periodic, 979 log-periodic, and 940 staggered range cells. Because the imposed increment is known exactly, the comparison measures branch recovery and recovered-shift error without treating any one estimator as truth. Width preservation is measured independently as the cell-by-cell difference between the translated and unshifted width estimates defined in Section~\ref{sec:real_data}.}

\begin{figure*}[!t]
    \centering
    \paperfig[width=0.90\textwidth]{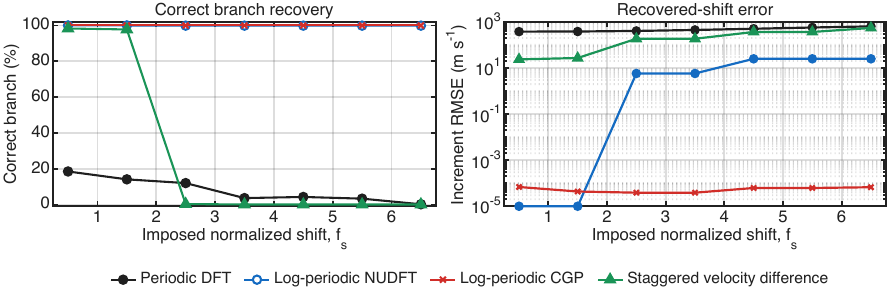}
    \caption{\revboth{Controlled Doppler translations of the measured complex echoes for SNR-gated range cells. The imposed truth is the normalized increment $f_s$ in \eqref{eq:injection_truth}; the left panel reports the branch-recovery rate $R_s$ in \eqref{eq:branch_recovery_rate}. The right panel reports the physical recovered-increment RMSE in \eqref{eq:injection_rmse} on a logarithmic scale; values below $10^{-5}$~\si{m.s^{-1}} are shown at that plotting floor. CGP on log-periodic data recovers every imposed branch, while periodic DFT and the velocity-difference estimator on staggered data deteriorate as the imposed shift increases. Periodic PSE has no branch-recovery curve because periodic sampling leaves the translated branches equivalent.}}
    \label{fig:measured_phase_injection}
\end{figure*}

\revboth{Fig.~\ref{fig:measured_phase_injection} shows that the log-periodic CGP recovers the correct branch in all 979 cells for every imposed shift. Its increment RMSE remains between $3.88\times10^{-5}$ and $6.85\times10^{-5}$~\si{m.s^{-1}}. The log-periodic NUDFT also performs well: it recovers every branch for $f_s=0.5$ and 1.5 and 99.90\% for the larger shifts. A single remote-branch error at each larger shift raises its RMSE to \SI{5.77}{m.s^{-1}} or \SI{24.95}{m.s^{-1}}, despite a median error of zero.}

\revboth{The velocity-difference estimator applied to the staggered data is reliable only for the first two shifts, with correct-branch rates of 98.30\% and 97.77\%. Its rate falls below 0.22\% from $f_s=2.5$ onward. The periodic DFT cannot determine the extended branch; its apparent recovery rate decreases from 18.43\% at $f_s=0.5$ to 0.10\% at $f_s=6.5$.}

\revtwo{The fitted widths remain effectively unchanged by the phase translation. For periodic PSE, the median cell-by-cell width change is zero for every imposed shift, and the middle half of the changes spans less than $2.2\times10^{-7}$~\si{m.s^{-1}}. The median changes for periodic DFT, log-periodic NUDFT, and log-periodic CGP remain within $3.3\times10^{-13}$~\si{m.s^{-1}} of zero. For the staggered velocity-difference estimator, the median change ranges from $-0.001$ to $-0.013$~\si{m.s^{-1}}, still small compared with the retrieved weather widths. Periodic PSE therefore retains the within-branch width but cannot identify the imposed extended branch because its mean is restricted to the principal periodic interval.}

\begin{figure*}[!t]
    \centering
    \paperfig[width=0.94\textwidth]{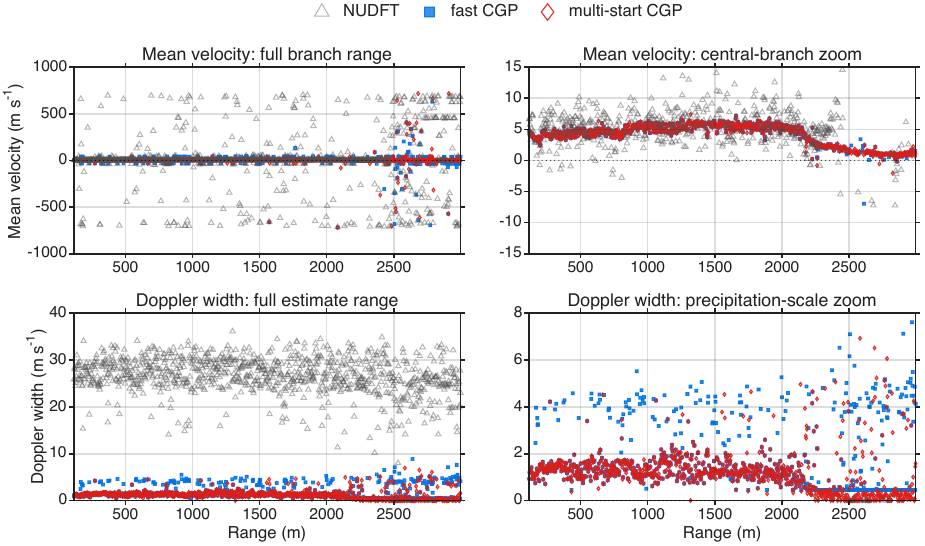}
    \caption{\revtwo{Range-resolved moment estimates for the log-periodic acquisition using $N=32$ samples. Open dark-gray triangles denote the log-periodic NUDFT, blue squares denote fast CGP, and open red diamonds denote the full multi-start CGP reference. The left column shows the full estimate range, including remote ambiguity branches and large NUDFT widths; the right column enlarges the precipitation-supported mean-velocity and Doppler-width ranges.}}
    \label{fig:stride2}
\end{figure*}

\begin{figure*}[!t]
    \centering
    \paperfig[width=0.92\textwidth]{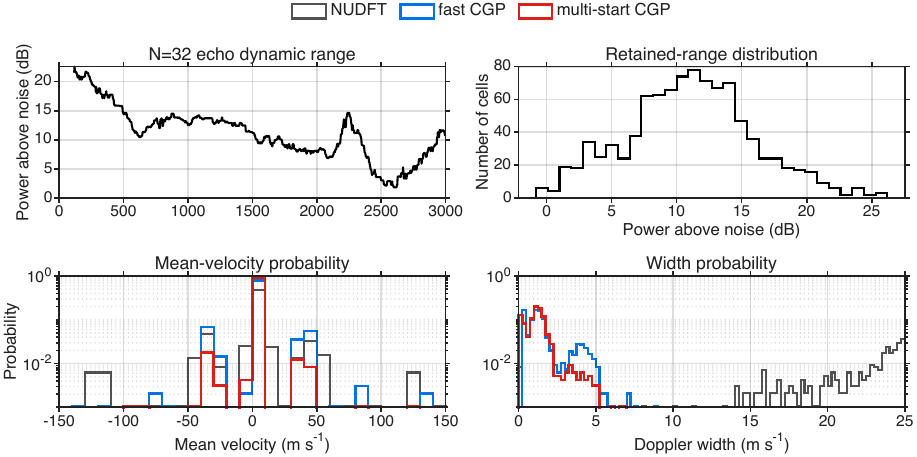}
    \caption{Power and moment distributions for the $N=32$ log-periodic comparison. The upper panels show power above the calibrated noise level over range and its cell distribution. The lower panels show probability-normalized mean-velocity and Doppler-width distributions; gray, blue, and red denote the \revtwo{log-periodic NUDFT}, fast CGP, and full multi-start CGP reference, respectively.}
    \label{fig:stride2_distributions}
\end{figure*}

The precipitation spectra in this event are narrow. Therefore, the \revtwo{log-periodic NUDFT} often selects the same branch as CGP when all 64 samples are used. Figs.~\ref{fig:stride2} and~\ref{fig:stride2_distributions} repeat the comparison using every second sample, reducing the record to $N=32$. The shorter record broadens the NUDFT response and produces more remote estimates and larger widths. The two CGP implementations remain close because both use the measured timestamps and the same covariance model. This agrees with \cite{Dash2024Counter-AliasingTrain}, where NUDFT artifacts increased with spectral width and shorter observations.

\revtwo{The present PSE formulation models a uniform finite observation window and is therefore applicable to the periodic acquisition. Its application to log-periodic timestamps would require an expected-periodogram model for the corresponding nonuniform observation window.}

\subsection{Range-Doppler Maps}
\begin{figure*}[!t]
    \centering
    \paperfig[width=0.89\textwidth]{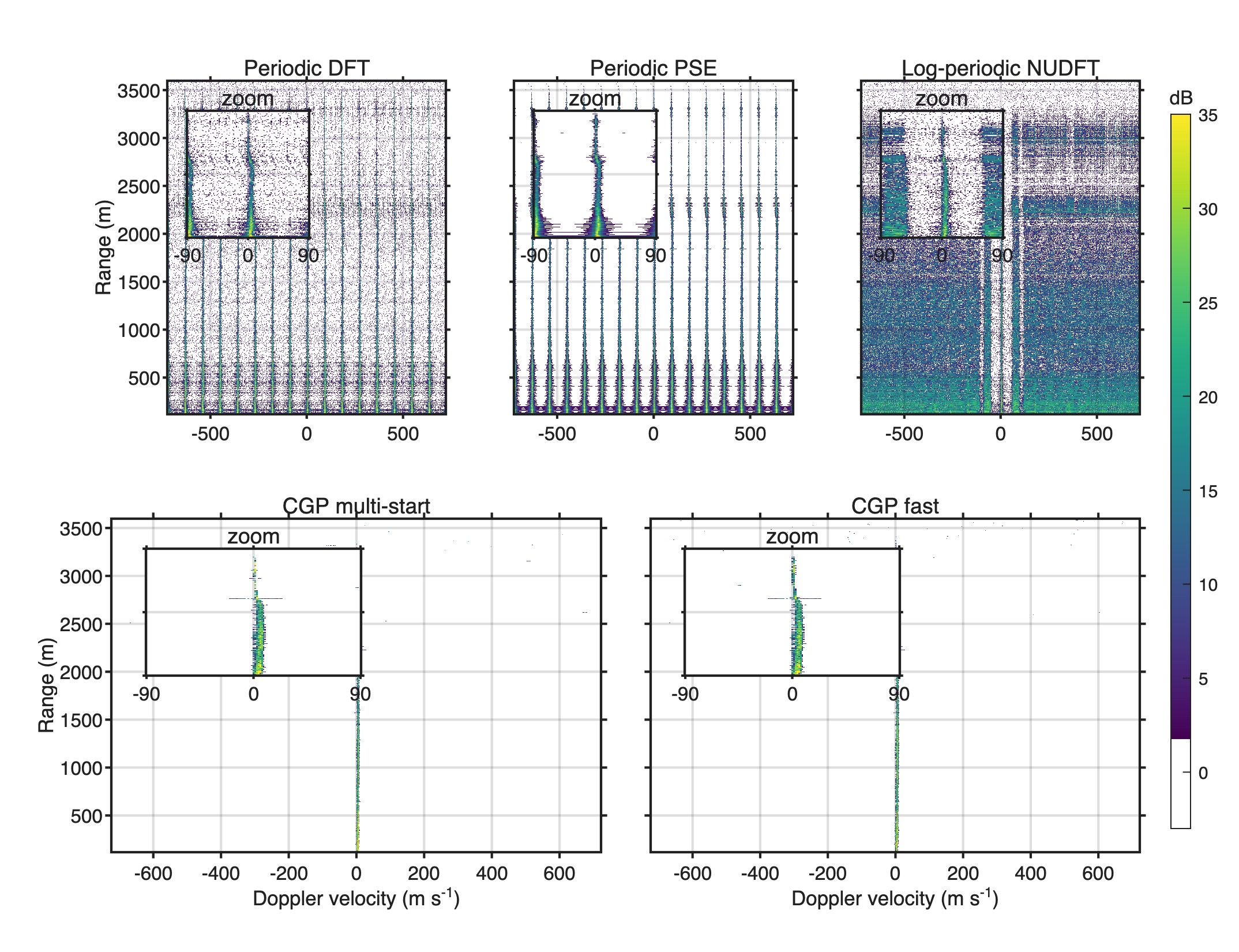}
    \caption{Range-Doppler maps for the compared reconstructions. Each panel shows the full extended velocity axis, with an inset zoom around the precipitation Doppler region. White denotes cells at or below the calibrated noise level; colored values show power above noise. \revtwo{The periodic PSE panel shows the fitted finite-window spectrum repeated over the unresolved periodic branches.}}
    \label{fig:rd_combined}
\end{figure*}

Fig.~\ref{fig:rd_combined} presents the reconstructed range-Doppler maps on the extended velocity axis, with an inset around the precipitation region in each panel. White indicates power at or below the calibrated noise level. The periodic DFT repeats the weather echo at every Nyquist interval. \revtwo{Periodic PSE narrows each copy but retains the repeated branches. The log-periodic NUDFT removes the exact repetition, although residual sidelobes remain.} The full and fast CGP maps are almost identical and place the precipitation power on one branch.

\section{Computational Complexity}
\label{sec:complexity}
Let $N$ be the number of slow-time samples, $M_f$ the number of Doppler grid points, $M_c$ the number of coarse CGP candidates, and $K$ the number of random starts used by the reference multi-start CGP estimator. A direct \revtwo{NUDFT} costs $\mathcal{O}(NM_f)$ per range cell. The staggered velocity-difference estimator is dominated by autocorrelation and branch selection and is approximately $\mathcal{O}(N+L)$, where $L$ is the number of ambiguity branches in the transfer function. The CGP likelihood requires a Cholesky factorization of the $N\times N$ covariance matrix, so one likelihood evaluation costs $\mathcal{O}(N^3)$. A multi-start CGP inversion costs approximately $\mathcal{O}(K I N^3)$, where $I$ is the number of local optimizer evaluations per start.

The fast CGP method costs
\begin{equation}
    \mathcal{O}(NM_f) + \mathcal{O}(M_cN^3) + \mathcal{O}(I_fN^3),
    \label{eq:fast_complexity}
\end{equation}
where $I_f$ is the number of evaluations in the final single local optimization. Since $N=64$ is small, the Cholesky factorization is not expensive by itself. The main practical gain is that the number of optimizer starts is no longer random or large. The candidate scoring is deterministic, parallel over candidates, and well suited to graphics processing unit (GPU) implementation because each likelihood evaluation has the same covariance structure and small dense linear algebra.

The prototype implementation is dominated by the number of local optimizations launched per range cell. A controlled single-cell benchmark was performed with $N=64$, the measured log-periodic time stamps, the same likelihood implementation, and ten independent synthetic weather realizations. The reference used 64 \texttt{patternsearch} starts, matching the full reference implementation, whereas the fast method included deterministic candidate scoring, both refinement stages, and one final \texttt{patternsearch} optimization. The median wall-clock times were \SI{3.01}{s} for the 64-start reference and \SI{0.338}{s} for the proposed method, corresponding to an $8.9\times$ speedup; both methods selected the correct basin in all ten trials. These timings are implementation and hardware dependent, but they quantify the scaling argument while including the overhead of the proposed basin search itself. Candidate scoring is independent across candidates and can be parallelized or transferred to a GPU.

\section{Discussion}
\label{sec:discussion}
\revboth{The moment-space ambiguity function gives the theoretical results a direct radar interpretation. Its central lobe is the neighborhood in which both Doppler mean and width remain locally resolvable. Remote sidelobes are competing mean-width pairs that produce a similar sampled covariance. A high sidelobe displaced mainly in mean Doppler represents a plausible wrong velocity branch; an extended response in the width direction indicates weak spectral-width resolution. Unlike a point-target ambiguity function, this representation shows how the finite width of a distributed weather echo changes the ambiguity pattern itself.}

\revboth{Clock precision determines the interval over which exact repetition can be avoided, but it does not guarantee that every spectrum in that interval is practically distinguishable. The moment-space maps connect these two statements. Periodic sampling produces repeated mean-Doppler sidelobes at the same width. Staggered sampling disrupts the simple repetition but retains strong structured sidelobes. The log-periodic pairwise lags suppress the remote sidelobes most clearly for narrow and moderate spectra. As the spectrum broadens, its covariance decays more rapidly with lag; the long-lag information that distinguishes remote branches is then lost, and the sidelobes of every timing law rise toward the main lobe. The false-detection transition near $\sigma_{fn}=0.16$ marks the onset of this information-limited regime for the tested design and SNR.}

\revone{The $d_2$ analysis makes the log-periodic sequence a tunable acquisition design rather than one fixed list of waiting times. Starting from the periodic limit, increasing $d_2$ lowers the strongest remote sidelobe for narrow and moderate spectra, but it also increases the dwell. The implemented value provides an appreciable sidelobe reduction without selecting the longest tested aperture. The preferred value should therefore be chosen from the expected spectral-width range and the available dwell, while recognizing that timing design alone cannot restore long-lag coherence for a very broad spectrum.}

\revtwo{The known-width comparison separates spectral-width estimation from ambiguity reduction. The DFT and NUDFT widths are broadened by the short observation window. PSE accounts for this window in its spectral likelihood, while CGP works directly with the covariance of the time samples. These parametric methods therefore recover narrow widths more accurately. Fast CGP closely follows the local CGP reference until the spectrum approaches the broad-width ambiguity region, showing that the proposed initialization usually finds the same likelihood basin.}

\revtwo{The theoretical PSE and CGP standard deviations decrease as SNR increases. Their difference is largest for narrow spectra, where the exact time-domain covariance retains information that is only approximated by the spectral likelihood. The measured data support the same interpretation: periodic PSE gives a median width of \SI{0.94}{m.s^{-1}}, compared with \SI{3.96}{m.s^{-1}} for the periodic DFT. PSE also preserves its width under phase translation, but it cannot identify the translated branch. Thus, the likelihood model determines the estimated width, whereas the timing law determines whether competing Doppler branches can be distinguished.}

The measured precipitation spectra are narrow over most of the analyzed range. With 64 samples, the \revtwo{log-periodic NUDFT} therefore often selects the same branch as CGP. With 32 samples, its sidelobes and width errors become more visible. Large velocities found in weak cells should not be interpreted as precipitation fall speeds; they indicate that the data in those cells cannot distinguish the main response from a remote sidelobe. The SNR gate removes these weak-cell selections from the quantitative comparison.

\revboth{The phase-injection result complements the sequential acquisition comparison by providing an exact Doppler increment within each measured record. It preserves the measured amplitudes, fluctuations, noise, and timing law while moving the complete spectrum onto a known branch. The perfect branch recovery of the log-periodic CGP over all tested shifts is therefore evidence of extended-domain statistical identifiability for these measured echoes. It is not a moving-target experiment: the injection does not introduce range migration or FMCW range-Doppler displacement, and the physical limits quantified in Section~\ref{sec:real_data} still apply.}

\revtwo{The staggered result also illustrates the difference between a formal and a practical ambiguity interval. Its high-precision timestamps imply a large hidden-grid interval, but the conventional velocity-difference estimator for staggered sampling uses two pulse-pair phases and a branch-selection transfer function rather than the full sampled covariance. A large clock-grid interval by itself therefore does not ensure reliable branch selection.}

\revone{The fast CGP procedure is an optimization strategy for the existing CGP MLE, not a separate statistical estimator. It uses deterministic candidates and local refinements to identify a promising likelihood basin before one final constrained optimization. The full and fast implementations consequently share the same covariance model, objective function, and final parameter bounds. Neither implementation uses continuity in range or a learned regression model; such information could be added for tracking, but it is not required for the single-cell results reported here.}

\revboth{The measured timing sequences are represented on a verified \SI{120}{MHz} clock, yet physical target motion limits the useful extent of the formal clock-grid interval. A target moving through a range cell during one dwell violates the stationary single-cell model, and FMCW Doppler also produces range displacement. The extended search interval should therefore be interpreted as the domain available to the sampled model, not as a claim that precipitation or other targets can be reconstructed at arbitrary velocity without accounting for migration and waveform coupling.}

\section{Conclusion}
\label{sec:conclusion}
\revboth{We develop and experimentally realize a Doppler counter-aliasing strategy in which only the waiting times between FMCW chirps change. The chirp duration, bandwidth, fast-time sampling, and range-processing chain remain unchanged. The acquisition can therefore alter the Doppler ambiguity pattern while retaining the long illumination and processing gain required for weather-radar range measurements.}

\revone{A central contribution is the moment-space ambiguity function for distributed Doppler spectra. By expressing expected likelihood similarity over candidate mean Doppler and spectral width, it identifies a main lobe around the true moment pair and sidelobes associated with competing velocity branches or width solutions. The function shows why log-periodic timing suppresses remote ambiguity for narrow and moderate spectra and why this protection decreases as broader spectra lose long-lag covariance.} \revboth{For the tested design, low false-detection rates are retained over 16 conventional periodic Nyquist intervals through $\sigma_{fn}=0.15$, before a rapid transition begins near $\sigma_{fn}=0.16$.} \revone{The accompanying $d_2$ analysis further shows that the sidelobe level can be traded against dwell through the curvature of the timing law.}

\revtwo{The known-width simulations over \SIlist{10;20;30}{dB} apply DFT and PSE to the periodic grid, NUDFT and the local and fast CGP estimators to the log-periodic grid, and the velocity-difference estimator to the staggered grid. The large widths returned by the uncorrected Fourier moments are dominated by finite-window broadening for the short dwells considered here. PSE and the staggered velocity-difference estimator reduce this bias, while the time-domain CGP likelihood gives the smallest width variability for the narrow cases; fast CGP closely follows the local exact-likelihood reference. Applying PSE to the measured periodic acquisition gives a median width of \SI{0.94}{m.s^{-1}} and preserves that width under controlled Doppler translations, while remaining unable to identify the translated periodic branch. These results separate estimator behavior from sampling design: parametric likelihoods provide narrow spectral-width estimates, whereas the aperiodic timing law controls the competing extended-domain ambiguity sidelobes.}

\revboth{The measured log-periodic acquisition demonstrates that the proposed timing law operates on PARSAX through high-precision trigger scheduling without modifying the transmitted chirp. Across 979 precipitation-supported range cells, the log-periodic CGP solution remains on the central branch over the extended search domain. Under known normalized Doppler shifts from 0.5 to 6.5, it recovers every branch in all 979 cells and preserves the fitted width. The $N=32$ comparison clarifies the benefit of covariance-based reconstruction as the nonuniform Fourier response broadens. The deterministic basin-selection procedure retains the CGP likelihood and final optimizer while reducing the median single-cell computation from \SI{3.01}{s} to \SI{0.338}{s} in the controlled benchmark.}

\revboth{Aperiodic slow-time triggering thus offers a practical route for weakening the usual coupling among illumination time, operational range, Doppler resolution, and unambiguous velocity in continuously transmitting radars. The present realization uses a Gaussian precipitation spectrum model; other target classes require an appropriate covariance or a parametric Doppler description. Within that scope, the results establish the acquisition principle, provide a moment-domain measure for designing its ambiguity response, and demonstrate fast extended-domain inversion on real weather-radar measurements.

The ambiguity function formulation in the moment space also gives us a powerful, compact tool for designing optimal sampling frameworks for generic user-defined problems in the future.}

\section*{Acknowledgment}
The authors very much appreciate the help provided by Thomas Epailly, and ing. F. van der Zwan during the experiments. This work is supported by the \textit{Phased Array Radar for Atmospheric Research} (PHARA) project from the Dutch Research Council (NWO) (Research Infrastructure (RI): national consortia; file number 175.2021.038). The timestamp sequences and analysis scripts are included in the 4TU-repository \cite{Dash_ASCEND_Dataset_2026}

\FloatBarrier
\bibliographystyle{IEEEtran}
\bibliography{references}

\begin{thebibliography}{10}
\providecommand{\url}[1]{#1}
\csname url@samestyle\endcsname
\providecommand{\newblock}{\relax}
\providecommand{\bibinfo}[2]{#2}
\providecommand{\BIBentrySTDinterwordspacing}{\spaceskip=0pt\relax}
\providecommand{\BIBentryALTinterwordstretchfactor}{4}
\providecommand{\BIBentryALTinterwordspacing}{\spaceskip=\fontdimen2\font plus
\BIBentryALTinterwordstretchfactor\fontdimen3\font minus
  \fontdimen4\font\relax}
\providecommand{\BIBforeignlanguage}[2]{{%
\expandafter\ifx\csname l@#1\endcsname\relax
\typeout{** WARNING: IEEEtran.bst: No hyphenation pattern has been}%
\typeout{** loaded for the language `#1'. Using the pattern for}%
\typeout{** the default language instead.}%
\else
\language=\csname l@#1\endcsname
\fi
#2}}
\providecommand{\BIBdecl}{\relax}
\BIBdecl

\bibitem{James2001}
C.~N. James and J.~Houze, ``{A real-time four-dimensional Doppler dealiasing
  scheme},'' \emph{Journal of Atmospheric and Oceanic Technology}, vol.~18,
  no.~10, pp. 1674--1683, 2001.

\bibitem{Zhang2006}
J.~Zhang and S.~Wang, ``{An automated 2D multipass Doppler radar velocity
  dealiasing scheme},'' \emph{Journal of Atmospheric and Oceanic Technology},
  vol.~23, no.~9, pp. 1239--1248, 2006.

\bibitem{Gao2004}
J.~Gao, K.~K. Droegemeier, J.~Gong, and Q.~Xu, ``{A method for retrieving mean
  horizontal wind profiles from single-Doppler radar observations contaminated
  by aliasing},'' \emph{Monthly Weather Review}, vol. 132, no.~6, pp.
  1399--1409, 2004.

\bibitem{Louf2020}
V.~Louf, A.~Protat, R.~C. Jackson, S.~M. Collis, and J.~Helmus, ``{Unravel: A
  robust modular velocity dealiasing technique for doppler radar},''
  \emph{Journal of Atmospheric and Oceanic Technology}, vol.~37, no.~5, pp.
  741--758, 2020.

\bibitem{UtkuPMCW}
N.~Petrov, U.~Kumbul, C.~S. Vaucher, and A.~Yarovoy, ``{Phase Modulated FMCW
  Waveforms and Receiver Structures for Automotive MIMO Radars},'' in
  \emph{2023 IEEE International Radar Conference (RADAR)}.\hskip 1em plus 0.5em
  minus 0.4em\relax IEEE, 2023, pp. 1--6.

\bibitem{Schvartzman2024}
D.~Schvartzman and R.~D. Palmer, ``{Doppler Velocity Recovery and Dealiasing
  Algorithm for Multi-PRT Scans in Weather Radars},'' \emph{IEEE Transactions
  on Geoscience and Remote Sensing}, vol.~62, pp. 1--14, 2024.

\bibitem{ImprovingNEXRADVelocityRetrievalsUsingMultiPRTScanswithRegressionProcessing}
\BIBentryALTinterwordspacing
J.~C. Hubbert, D.~Schvartzman, and U.~Romatschke, ``{Improving NEXRAD Velocity
  Retrievals Using Multi-PRT Scans with Regression Processing},'' \emph{Journal
  of Atmospheric and Oceanic Technology}, vol.~43, no.~4, pp. 475--488, 2026.
  [Online]. Available:
  \url{https://journals.ametsoc.org/view/journals/atot/43/4/JTECH-D-25-0059.1.xml}
\BIBentrySTDinterwordspacing

\bibitem{Dash2024Counter-AliasingTrain}
T.~Dash, H.~Driessen, O.~A. Krasnov, and A.~Yarovoy, ``{Counter-Aliasing Is
  Better Than De-Aliasing: Application to Doppler Weather Radar With Aperiodic
  Pulse Train},'' \emph{IEEE Transactions on Geoscience and Remote Sensing},
  vol.~62, pp. 1--17, 2024.

\bibitem{Doviak1993}
R.~J. Doviak and D.~S. Zrni{\'c}, \emph{{Doppler Radar and Weather
  Observations}}, 2nd~ed.\hskip 1em plus 0.5em minus 0.4em\relax San Diego, CA,
  USA: Academic Press, 1993.

\bibitem{Zrnic1975VelocityRadar}
D.~S. Zrni{\'c} and R.~J. Doviak, ``{Velocity Spectra of Vortices Scanned with
  a Pulse-Doppler Radar},'' \emph{Journal of Applied Meteorology}, vol.~14,
  no.~8, pp. 1531--1539, 12 1975.

\bibitem{Zrnic1977}
D.~S. Zrni{\'c}, ``{Spectral Moment Estimates from Correlated Pulse Pairs},''
  \emph{IEEE Transactions on Aerospace and Electronic Systems}, vol. AES-13,
  no.~4, pp. 344--354, 1977.

\bibitem{Sachidananda2000ClutterPRT}
M.~Sachidananda and D.~S. Zrni{\'c}, ``{Clutter filtering and spectral moment
  estimation for Doppler weather radars using staggered pulse repetition time
  (PRT)},'' \emph{Journal of Atmospheric and Oceanic Technology}, vol.~17,
  no.~3, pp. 323--331, 2000.

\bibitem{Torres2004DesignWSR-88D}
S.~M. Torres, Y.~F. Dubel, and D.~S. Zrni{\'c}, ``{Design, implementation, and
  demonstration of a staggered PRT algorithm for the WSR-88D},'' \emph{Journal
  of Atmospheric and Oceanic Technology}, vol.~21, no.~9, pp. 1389--1399, 2004.

\bibitem{Sachidananda2003UnambiguousTechnique}
M.~Sachidananda and D.~S. Zrni{\'c}, ``{Unambiguous range extension by overlay
  resolution in staggered PRT technique},'' \emph{Journal of Atmospheric and
  Oceanic Technology}, vol.~20, no.~5, pp. 673--684, 2003.

\bibitem{Moisseev2008ClutterWaveforms}
D.~N. Moisseev, C.~M. Nguyen, and V.~Chandrasekar, ``{Clutter suppression for
  staggered PRT waveforms},'' \emph{Journal of Atmospheric and Oceanic
  Technology}, vol.~25, no.~12, pp. 2209--2218, 2008.

\bibitem{Warde2017Staggered-PRTFilter}
D.~A. Warde and S.~M. Torres, ``{Staggered-PRT Sequences for Doppler Weather
  Radars. Part II: Ground Clutter Mitigation on the NEXRAD Network Using the
  CLEAN-AP Filter},'' \emph{Journal of Atmospheric and Oceanic Technology},
  vol.~34, no.~3, pp. 703--716, 2017.

\bibitem{Torres2017}
\BIBentryALTinterwordspacing
S.~M. Torres and D.~A. Warde, ``{Staggered-PRT Sequences for Doppler Weather
  Radars. Part I: Spectral Analysis Using the Autocorrelation Spectral
  Density},'' \emph{Journal of Atmospheric and Oceanic Technology}, vol.~34,
  no.~1, pp. 51--63, 1 2017. [Online]. Available:
  \url{https://journals.ametsoc.org/view/journals/atot/34/1/jtech-d-16-0071.1.xml}
\BIBentrySTDinterwordspacing

\bibitem{Scargle1982StudiesData}
J.~D. Scargle, ``{Studies in astronomical time series analysis. II -
  Statistical aspects of spectral analysis of unevenly spaced data},''
  \emph{Astrophysical Journal}, vol. 263, pp. 835--853, 1982.

\bibitem{VanderPlas2018UnderstandingPeriodogram}
\BIBentryALTinterwordspacing
J.~T. VanderPlas, ``{Understanding the Lomb–Scargle Periodogram},'' \emph{The
  Astrophysical Journal Supplement Series}, vol. 236, no.~1, p.~16, 2018.
  [Online]. Available: \url{http://dx.doi.org/10.3847/1538-4365/aab766}
\BIBentrySTDinterwordspacing

\bibitem{Doerry2017RadarPRF}
\BIBentryALTinterwordspacing
A.~W. Doerry, ``{Radar Doppler processing with nonuniform PRF},'' Sandia
  National Laboratories Albuquerque, Albuquerque, New Mexico, Tech. Rep.
  SAND2017-7851, 2017. [Online]. Available:
  \url{https://www.osti.gov/servlets/purl/1373645}
\BIBentrySTDinterwordspacing

\bibitem{Gaster1977TheTransform}
M.~Gaster and J.~Roberts, ``{The spectral analysis of randomly sampled records
  by a direct transform},'' \emph{Proc. Roy. Soc. A.}, vol. 354, no. 1676, pp.
  27--58, 1977.

\bibitem{S.A.KSyedMohamed2024ApplicationSpectrum}
{S.A.K Syed Mohamed}, T.~Dash, O.~A. Krasnov, J.~Bout, R.~v.~d. Meer, and A.~G.
  Yarovoy, ``{Application of non-uniform sampling to avoid aliasing in the
  precipitation Doppler spectrum},'' \emph{2024 International Radar Symposium
  (IRS)}, pp. 216--221, 2024.

\bibitem{Palmer1991}
R.~D. Palmer, J.~R. Cruz, and D.~S. Zrni{\'c}, ``{Enhanced autoregressive
  moving average spectral estimation applied to the measurement of Doppler
  spectral width},'' \emph{IEEE Transactions on Geoscience and Remote Sensing},
  vol.~29, no.~3, pp. 358--368, 5 1991.

\bibitem{Pinsky2011}
M.~Pinsky, J.~Figueras I~Ventura, T.~Otto, A.~Sterkin, A.~Khain, and H.~W.
  Russchenberg, ``{Application of a simple adaptive estimator for an
  atmospheric Doppler Radar},'' \emph{IEEE Transactions on Geoscience and
  Remote Sensing}, vol.~49, no. 1 PART 1, pp. 115--127, 2011.

\bibitem{Warde2014TheAnalysis}
D.~A. Warde and S.~M. Torres, ``{The autocorrelation spectral density for
  doppler-weather-radar signal analysis},'' \emph{IEEE Transactions on
  Geoscience and Remote Sensing}, vol.~52, no.~1, pp. 508--518, 2014.

\bibitem{Thomson1982SpectrumAnalysis}
D.~J. Thomson, ``{Spectrum Estimation and Harmonic Analysis},''
  \emph{Proceedings of the IEEE}, vol.~70, no.~9, pp. 1055--1096, 1982.

\bibitem{Das2021SpectralApproach}
S.~Das, S.~Subba~Rao, and J.~Yang, ``{Spectral methods for small sample time
  series: A complete periodogram approach},'' \emph{Journal of Time Series
  Analysis}, vol.~42, no. 5-6, pp. 597--621, 2021.

\bibitem{Ariananda2012CompressiveEstimation}
D.~D. Ariananda and G.~Leus, ``{Compressive wideband power spectrum
  estimation},'' \emph{IEEE Transactions on Signal Processing}, vol.~60, no.~9,
  pp. 4775--4789, 2012.

\bibitem{Purnamasari2021}
R.~Purnamasari, A.~B. Suksmono, I.~Zakia, and I.~J.~M. Edward, ``{Compressive
  Sampling of Polarimetric Doppler Weather Radar Processing Via Inverse Fast
  Fourier Transform},'' \emph{IEEE Journal of Selected Topics in Applied Earth
  Observations and Remote Sensing}, vol.~14, pp. 5269--5284, 2021.

\bibitem{Dash2024DopplerModel}
T.~Dash, H.~Driessen, O.~A. Krasnov, and A.~Yarovoy, ``{Doppler Spectrum
  Parameter Estimation for Weather Radar Echoes Using a Parametric
  Semianalytical Model},'' \emph{IEEE Transactions on Geoscience and Remote
  Sensing}, vol.~62, pp. 1--18, 2024.

\bibitem{Tobar2018BayesianEstimation}
F.~Tobar, ``{Bayesian nonparametric spectral estimation},'' \emph{Advances in
  Neural Information Processing Systems}, vol. 2018-Decem, no. NeurIPS, pp.
  10\,127--10\,137, 2018.

\bibitem{Dash2023PrecipitationPrior}
T.~K. Dash, H.~Driessen, O.~A. Krasnov, and A.~Yarovoy, ``{Precipitation
  Doppler Spectrum Reconstruction with Gaussian Process Prior},'' \emph{2023
  IEEE Conference on Antenna Measurements and Applications (CAMA)}, pp.
  909--914, 2023.

\bibitem{Boloix-Tortosa2018ComplexRegression}
R.~Boloix-Tortosa, J.~J. Murillo-Fuentes, F.~J. Payan-Somet, and F.~Perez-Cruz,
  ``{Complex Gaussian Processes for Regression},'' \emph{IEEE Transactions on
  Neural Networks and Learning Systems}, vol.~29, no.~11, pp. 5499--5511, 2018.

\bibitem{Kullback1951}
S.~Kullback and R.~A. Leibler, ``{On Information and Sufficiency},'' \emph{The
  Annals of Mathematical Statistics}, vol.~22, no.~1, pp. 79--86, 1951.

\bibitem{CoverThomas2006}
T.~M. Cover and J.~A. Thomas, \emph{{Elements of Information Theory}},
  2nd~ed.\hskip 1em plus 0.5em minus 0.4em\relax Hoboken, NJ, USA:
  Wiley-Interscience, 2006.

\bibitem{Mathworks2024OptimizationR2024a}
\BIBentryALTinterwordspacing
{Mathworks}, ``{Optimization Toolbox™ User's Guide R2024a},'' 2024. [Online].
  Available: \url{www.mathworks.com}
\BIBentrySTDinterwordspacing

\bibitem{Krasnov2010PARSAX}
O.~A. Krasnov, G.~Babur, Z.~Wang, L.~P. Ligthart, and W.~F. van~der Zwan,
  ``Basics and first experiments demonstrating isolation improvements in the
  agile polarimetric {FM-CW} radar---{PARSAX},'' \emph{International Journal of
  Microwave and Wireless Technologies}, vol.~2, no. 3--4, pp. 419--428, 2010.

\bibitem{Levin1965PowerEstimation}
M.~J. Levin, ``{Power Spectrum Parameter Estimation},'' \emph{IEEE Transactions
  on Information Theory}, vol.~11, no.~1, pp. 100--107, 1965.

\bibitem{Dash_ASCEND_Dataset_2026}
T.~Dash, S.~A.~K. Syed~Mohamed, O.~Krasnov, A.~Yarovoy, T.~Epailly, and
  F.~van~der Zwan, ``{ASCEND}: Aperiodic sampling for counter-aliasing of
  extended-nyquist doppler spectra,'' 2026, unpublished dataset.

\end{thebibliography}

\end{document}